%% file: main.tex
\documentclass[10pt,journal,final,letterpaper,oneside,twocolumn]{IEEEtran}
\usepackage{amsmath,amsfonts}
\usepackage{hyperref}
\usepackage{mathtools}
\usepackage{tcolorbox}
\usepackage{color}
\usepackage{theorem}
\usepackage{times,epsfig}
\usepackage{amssymb}
\usepackage{subfigure}
\usepackage{cite}
\usepackage{cases}
\usepackage{siunitx}
\usepackage{bm}

\usepackage{algorithmic}
\usepackage{algorithm}
\input{my_sections.tex}

\usepackage{tikz}
\usetikzlibrary{shapes,arrows}
\input{mysymbol.sty}
\input{figures/tikz_styles}

\DeclareMathOperator{\Tr}{Trace}
\DeclareMathAlphabet\mathbfcal{OMS}{cmsy}{b}{n}

\newtheorem{mytheorem}{\bf Theorem}
\newtheorem{mydefinition}{\bf Definition}

\newtheorem{myproposition}{\bf Proposition}
\newtheorem{remark}{\bf Remark}

\newtcolorbox{myblockt}[1]{colback=urblue!5!white,
	colframe=urblue,fonttitle=\bfseries,
	title=#1}

\newtcolorbox{myblock}{colback=urblue!5!white,
	colframe=urblue,fonttitle=\bfseries}

\title{Deep Graph Unfolding for Beamforming in MU-MIMO Interference Networks}

\author{\IEEEauthorblockN{Arindam Chowdhury, Gunjan Verma, Ananthram Swami, and Santiago Segarra}
\thanks{Research was sponsored by the Army Research Office and was accomplished under         Cooperative Agreement Number W911NF-19-2-0269. 
        The views and conclusions contained in this document are those of the authors and should not be interpreted as representing the official policies, either expressed or implied, of the Army Research Office or the U.S. Government. 
        The U.S. Government is authorized to reproduce and distribute reprints for Government purposes notwithstanding any copyright notation herein. Preliminary results were presented in~\cite{chowdhury2021ml}.
        \newline
        A. Chowdhury and S. Segarra are with the Dept. of ECE, Rice University.
	G. Verma, and A. Swami are with the US Army’s DEVCOM Army Research Laboratory.
        \newline
	Emails:  \{ac131, segarra\}@rice.edu, \hspace{1mm} \{gunjan.verma, ananthram.swami\}.civ@army.mil.}
}

\begin{document}
\maketitle

\begin{abstract}
    We develop an efficient and near-optimal solution for
    beamforming in multi-user multiple-input-multiple-output single-hop wireless ad-hoc interference networks.
    Inspired by the weighted minimum mean squared error (WMMSE) method, a classical approach to solving this problem, and the principle of algorithm unfolding, we present unfolded WMMSE (UWMMSE) for MU-MIMO. 
    This method learns a parameterized functional transformation of key WMMSE parameters using graph neural networks (GNNs), where the channel and interference components of a wireless network constitute the underlying graph. 
    These GNNs are trained through gradient descent on a network utility metric using multiple instances of the beamforming problem.
    Comprehensive experimental analyses illustrate the superiority of UWMMSE over the classical WMMSE and state-of-the-art learning-based methods in terms of performance, generalizability, and robustness.
\end{abstract}


\section{Introduction}

Multi-user multi-input-multi-output (MU-MIMO)~\cite{heath2018foundations,bjornson2017massive} systems have become increasingly useful in the context of multi-antenna beamforming~\cite{bjornson2015optimal,ngo2013energy} in both multi-cell~\cite{razaviyayn2013linear} and ad-hoc~\cite{luo2008dynamic} wireless network scenarios. 
They are especially beneficial for increasing spectral efficiency and improving effective network capacity to meet the high quality-of-service (QoS) requirements of modern wireless systems~\cite{liang2019deep}.
The task of multi-antenna beamforming is particularly challenging for wireless ad-hoc networks (WANETs) wherein the transceivers may operate under strict power constraints. 
For example, a mission-specific military deployment might use multiple hand-held devices with limited battery life to constitute a tactical WANET.
Moreover, these deployments can be in various topological, environmental, and weather conditions, giving rise to varying fading effects and path loss.
Also, the devices in the network may suffer from interference with each other, posing further challenges towards maintaining the required QoS.
Broadly then, the key task of beamforming in a MU-MIMO WANET involves managing the channel and interference conditions in the wireless network, to obtain beams that achieve a reasonably high value of a given QoS metric without violating the overall power constraint.

The beamforming task can be formalized as an optimization problem involving a network utility function -- e.g., sum-rate, mean-rate, or harmonic-rate -- as the objective with resource constraints, such as maximum available power, at each transceiver.
Optimization problems of this form have been shown to be non-convex and NP-hard,~\cite{luo2008dynamic,liu2011coordinated,razaviyayn2013linear} and therefore lack closed-form solutions.
In the absence of methods to generate exact solutions to the beamforming problem, multiple classical approaches have been proposed that try to obtain approximate solutions. 
For the most common case of sum-rate maximization, the following broad categories of methods have been commonly employed in the last few decades: Lagrangian dual decomposition~\cite{yu2006dual,hayashi2009spectrum}, successive convex approximation~\cite{papandriopoulos2009scale}, interference pricing~\cite{wang2008price} and weighted minimum mean squared error (WMMSE) minimization \cite{shi2011iteratively}.

WMMSE is a popular algorithm for beamforming in both multi-cell and ad-hoc wireless networks. 
It offers closed-form, iterative update rules to solve a surrogate optimization problem that has been shown to have identical local optima as the sum-rate optimization problem~\cite{shi2011iteratively}.
In spite of being an effective classical solution, it has several drawbacks. First, it is computationally complex on account of several matrix inversion and eigendecomposition operations in each iteration.
Additionally, WMMSE has to be applied from scratch each time a new channel condition is presented, creating a significant lag in obtaining successive beamformers.
Finally, the solution offered by WMMSE is near-optimal at best, as it can only generate a local optimum of the sum-rate optimization problem. 

Deep learning methods have also been proposed to solve the challenging beamforming problem.
More specifically, these neural models take a channel state information (CSI) tensor -- a consolidated structure containing channel and interference measures between pairs of indexed nodes -- as input and generate the corresponding beamformer for all the transmitters in the network.
Several neural architectures of varying complexity -- multi-layer perceptrons (MLP)~\cite{sun2018learning,sun2021learning}, convolutional neural networks (CNN)~\cite{lee2018deep}, recurrent neural networks (RNN)~\cite{deng2019application}, and even graph neural networks (GNN)~\cite{eisen2019learning,eisen2020optimal} -- have been applied to this end.
A major advantage of these methods lies in their low feedforward computational complexity.
Moreover, these methods generate the beamformers in parallel for all nodes in the network, making them applicable to large-scale networks.
Nevertheless, these methods fail to generalize to channel conditions that were unseen at training.
Further, they often fail to be competitive with WMMSE since they lack the domain-specific information encoded in the structured WMMSE updates. 

To address these drawbacks, a class of hybrid algorithms~\cite{hu2020iterative,pellaco2020deep,schynol2022coordinated,chowdhury2021unfolding,chowdhury2021ml} that combine the classical update structure of WMMSE with the fast inference capabilities of neural models have been proposed. 
This is achieved through the learning paradigm of algorithm unfolding~\cite{monga2019algorithm,balatsoukas2019deep,liu2019deep,farsad2020data}.
Algorithm unfolding decouples the update steps of an iterative algorithm to create a cascade of hybrid layers that preserve the original update structure but introduces one or more learnable parameters from data.
This form of domain-inspired learning has been extremely popular and effective in several application areas, including but not limited to non-negative matrix factorization~\cite{hershey2014deep}, iterative soft thresholding~\cite{gregor2010learning}, semantic image segmentation~\cite{liu2017deep}, blind deblurring~\cite{li2019algorithm}, clutter suppression~\cite{solomon2019deep}, particle filtering~\cite{gama2023unsupervised}, symbol detection~\cite{zilberstein2022detection}, link scheduling~\cite{zhao2022link}, energy-aware power allocation~\cite{li2022graph}, and beamforming in wireless networks~\cite{hu2020iterative,pellaco2020deep,schynol2022coordinated}.
These algorithms use various neural layers to learn one or more parameters of the iterative algorithm being unfolded or to approximate certain computational steps in the algorithm to reduce complexity and speed-up processing.

In this work, we propose an unfolding solution of WMMSE wherein we combine two separate classes of neural architectures with very specific advantages -- a ReLU MLP that enforces a parametric functional transformation on a specific WMMSE variable -- in a manner elaborated further in Section~\ref{Ss:unfold}, specifically in~\eqref{E:learn_w}, -- and a GNN which learns the parameters of the MLP by leveraging the underlying graph structure of the wireless network.
While both MLP~\cite{sun2018learning} and GNN~\cite{schynol2022coordinated} are common generic architectures, the specific form of unfolding proposed in this paper is entirely novel. 
The proposed scheme relies on the universal approximation property of MLPs to allow the unfolded variable to learn its own functional transformation -- necessary for better convergence -- without enforcing any prior structure to it.
At the same time, the use of GNNs to learn the transformation parameters allows the model to incorporate the connectivity information embedded in the wireless network that cannot be otherwise extracted for arbitrary network topologies by a general-purpose MLP.  

\noindent\textbf{Contribution.} 
Following are the three main contributions of this work:\\
i) We propose a hybrid algorithm, namely UWMMSE -- by unfolding WMMSE -- for beamforming in multi-user multi-input-multi-output wireless ad-hoc networks -- where the parameters of a functional transformation are learned via a GNN -- along with provisions for its distributed implementation. Further, we emphasize on obtaining a fast and lightweight model by extensive parameter-sharing to reduce model complexity.\\
ii) We present a theoretical analysis of the proposed model in terms of a necessary condition that the learnable transformation must satisfy to enable effective learning; [ref. Theorem~\ref{T:convergence_necessary}]. Additionally, we establish permutation equivariance of the proposed model; [ref. Proposition~\ref{P:equiv}].
\\
iii) We provide comprehensive experimental analyses to illustrate the empirical superiority of the proposed method. Firstly, we present a comparison with state-of-the-art connectionist methods. We then demonstrate the generalizability of the proposed model to unseen network sizes and its robustness to out-of-distribution inputs. 


\noindent\textbf{Notation.}
$[\mathbfcal{X}]_{ij\dots}$, $[\mathbf{X}]_{ij}$, and $[\mathbf{x}]_{i}$ denote the entries of a multi-dimensional tensor $\mathbfcal{X}$, a matrix $\mathbf{X}$, and a vector $\mathbf{x}$. 
The generic subindex $:$ denotes a whole dimension, e.g., row $i$ of matrix $\mathbf{X}$ is denoted as $[\mathbf{X}]_{i \,:}$.
$\mathbb{E}(\cdot)$ is the expectation operator while $(\cdot)^{H}$ represents conjugate transpose.
The all-zeros and all-ones tensors are denoted by $\mathbf{0}$ and $\mathbf{1}$, respectively, where the dimensions are clear from context.
The $z\times z$ identity matrix is represented by $\mathbf{I}_z$.
The diagonal matrix $\diag(\bbX)$ stores the diagonal elements of $\bbX$. 
The zero-mean complex-normal distribution is denoted by $\ccalC\ccalN(0, \sigma^2\mathbf{I})$.

\section{System Model and Problem Formulation}~\label{S:sys}

Our system is a single-hop MU-MIMO WANET with $M$ distinct transmitter-receiver pairs. 
Each transmitter, having $T$ antennas, transmits independent data signals to a corresponding receiver equipped with $R$ antennas. 
Let $\mathbf{V}_{i} \in \mathbb{C}^{T\times d}$ denote the beamformer that transmitter $i$ uses to transmit a signal $\mathbf{x}_i \in \mathbb{C}^{d}$, where $\mathbb{E}[\mathbf{x}_i\mathbf{x}_i^H] = \mathbf{I}_d$,  to its assigned receiver $r(i)$. 
Assuming a linear channel model, the signal $\mathbf{y}_i \in \mathbb{C}^{R}$, received at $r(i)$ is of the form
\begin{equation}\label{E:trans_model}
    \mathbf{y}_i = \underbrace{\mathbf{H}_{ii}\mathbf{V}_i\mathbf{x}_i}_{desired\ signal} + \underbrace{\sum_{\substack{j=1 \, | \,  j\neq i}}^M \mathbf{H}_{ij}\mathbf{V}_j\mathbf{x}_j + \mathbf{n}_i}_{interference\ plus\ noise}, \quad \text{for all } i,
\end{equation}
where $\mathbf{n}_i \in \mathbb{C}^{R}$ denotes independent additive white gaussian noise sampled from $\mathcal{CN}(0,\sigma^2\mathbf{I}_R)$. 
Here, $\mathbf{H}_{ii} \in \mathbb{C}^{R\times T}$ represents the communication channel between transmitter $i$ and its assigned receiver $r(i)$ while $\mathbf{H}_{ij} \in \mathbb{C}^{R\times T}$ for all $j \neq i$ represents the interference between $r(i)$ and all other transmitters $j$.
Finally, the transmitted signal is estimated at $r(i)$ using a receiver-beamformer $\mathbf{U}_{i} \in \mathbb{C}^{R\times d}$, to obtain $\Hat{\mathbf{x}}_i = \mathbf{U}_i^H \mathbf{y}_i$ for all $i \in {\{1, \dots, M\}}$.

If we define the channel state information (CSI) tensor $\mathbfcal{H} \in \mathbb{C}^{M\times M\times R\times T}$ such that $[\mathbfcal{H}]_{ij::} = \mathbf{H}_{ij}$ and the transmitter-beamformer tensor $\mathbfcal{V} \in \mathbb{C}^{M\times T\times d}$ such that $[\mathbfcal{V}]_{i::} = \mathbf{V}_{i}$, then for every user $i$, assuming perfect knowledge of the CSI matrices, its achievable rate~\cite{shi2011iteratively} is given by
\begin{align}\label{E:data_rate}
    c_i(\mathbfcal{V},\mathbfcal{H}) &= \log_2 \det ( \mathbf{I}_R + \mathbf{H}_{ii}^{}\mathbf{V}_{i}^{} \mathbf{V}_{i}^{H} \mathbf{H}_{ii}^{H} 
    ( \sigma^2\mathbf{I}_R + \nonumber \\ &\quad\quad\quad\quad\quad\quad\quad\quad \sum_{j\neq i} \mathbf{H}_{ij}^{}\mathbf{V}_{j}^{} \mathbf{V}_{j}^{H} \mathbf{H}_{ij}^{H} )^{-1} ) 
\end{align}

Our objective is to determine the $\mathbfcal{V}$ that maximizes the sum-rate of the whole network, 
\begin{align}\label{E:optimization_problem}
		 &\quad \quad \max_{\mathbfcal{V}} \,\, \sum_{i=1}^M \alpha_i c_i(\mathbfcal{V}, \mathbfcal{H}) \quad  \\& \text{s.t.} \,\,\,\,\,  \Tr\left(\mathbf{V}_i^{}\mathbf{V}_i^{H}\right) \leq P_{\mathrm{max}}, \,\,\,\, \forall \,\, i, \nonumber
\end{align}
where $\alpha_i \in \mathbb{R}$ represents the priority of the transceiver pair $i$ and the maximum power available uniformly to each transmitter is denoted by $P_{\max} \in \mathbb{R}$.
Henceforth, for simplicity, we focus on the case where every user is given the same $\alpha_i=1$ in the objective. 

The optimization problem in~(\ref{E:optimization_problem}) is non-convex and NP-hard~\cite{luo2008dynamic,hong_2014_signal}. 
A standard approach to solving this problem is to reformulate it as a constrained weighted-minimum-mean-square-error (WMMSE) optimization~\cite{shi2011iteratively}.
Specifically, introducing the receiver-weight tensor $\hat{\mathbfcal{W}} \in \mathbb{C}^{M\times d\times d}$ and receiver-beamformer tensor $\mathbfcal{U} \in \mathbb{C}^{M\times R\times d}$ the problem can be defined as 
\begin{align}\label{E:problem_reformulation}
&\min_{\mathbfcal{\hat{W},U,V}} \sum_{i=1}^M (\Tr(\hat{\mathbf{W}}_i \mathbf{E}_i) - \log \det \hat{\mathbf{W}}_i ),\\
& \text{s.t.} \,\,\,\,\,  \Tr\left(\mathbf{V}_i^{}\mathbf{V}_i^{H}\right) \leq P_{\mathrm{max}}, \,\,\,\, \forall \,\, i, \nonumber
\end{align}
where $\hat{\mathbf{W}}_i = [\hat{\mathbfcal{W}}]_{i::} \succeq 0$ is a weight matrix at receiver $r(i)$ while $\mathbf{E}_i \in \mathbb{C}^{d\times d}$ is the mean squared error between transmitted and received signals~\cite{shi2011iteratively}. 
%

The optimization problem~\eqref{E:problem_reformulation} is equivalent to~\eqref{E:optimization_problem} as shown in~\cite[Thm. 3]{shi2011iteratively} -- the variable $\mathbfcal{V}^*$ in the global optimal solution $\{\hat{\mathbfcal{W}}^*, \mathbfcal{U}^*, \mathbfcal{V}^*\}$ of the former is the same as the optimal transmitter-beamformer $\mathbfcal{V}^*$ in the latter. 
Moreover, the problem in~\eqref{E:problem_reformulation} is \emph{tri-convex}, i.e., fixing any two variables renders the objective function convex in the third variables.
This makes~\eqref{E:problem_reformulation} amenable to a block-coordinate-descent (BCD) based solution. 

In spite of the tri-convexity property that results in a tractable closed-form solution, WMMSE performance is limited by its cumbersome iterations that are composed of expensive computational steps like matrix inversion, eigendecomposition, and bisection search (per-iteration complexity of WMMSE scales as $\mathcal{O}(M^2)$, where $M$ is the number of users)~\cite{chowdhury2021unfolding}. 
Naturally, WMMSE tends to be time-consuming depending on the size and complexity of the wireless network, making it particularly ineffective for fast-changing channels.
Moreover, while WMMSE can achieve a near-optimal solution, it  can only solve a single instance of~\eqref{E:optimization_problem} for a given $\mathbfcal{H}$. In case there are multiple CSI tensors, say $\{\mathbfcal{H}_i\}_{i=1}^{n}$, to be processed -- e.g., in a scenario wherein multiple wireless sub-networks are being optimized over by a centralized optimizing agent --, WMMSE has to be repeated from scratch independently for each of the $n$ instances.  

From a practical standpoint, it is desirable to have a mechanism for \emph{fast}, \emph{efficient} and \emph{interpretable} processing of a set of independent CSI tensors. 
We propose to achieve this through a GNN-based unfolded algorithm. More specifically, we leverage the near-optimal solution provided by the iterations of the classical WMMSE method and enhance it with the expressivity and computational efficiency of trained graph neural models.

\section{Unfolded-WMMSE for MU-MIMO}\label{S:uwmmse}
Since WMMSE is composed of computationally expensive iterations, we reduce the computations while preserving the update structure by truncating the number of iterations and then compensating for the reduced iterations using data-driven neural modules.

\subsection{Designing the unfolded architecture}\label{Ss:unfold}

We define a $K$-layered parametric function $\Lambda( \cdot ; \Theta): \mathbb{C}^{M\times M\times R\times T} \to \mathbb{C}^{M\times T\times d}$ where $\Theta$ is a set of trainable parameters and $\mathbfcal{V}^{(K)} = \Lambda( \mathbfcal{H}; \Theta)$ approximates the solution to~\eqref{E:optimization_problem} for a given CSI tensor $\mathbfcal{H}$. 
The layers in $\Lambda$ are hybrid structures designed using the WMMSE updates~\cite{shi2011iteratively}, augmented by a learnable transformation $\Phi$ to accelerate convergence. 
More specifically, by setting the initial beamformers $[\mathbfcal{V}^{(0)}]_{i::} = \mathbf{V}^{(0)}_i = \mathrm{v}^{}_{\mathrm{init}}\mathbf{1}_{T \times d} \,$ where $\mathrm{v}^{}_{\mathrm{init}} = \sqrt{\frac{P_{\max}}{2Td}}(1+\sqrt{-1})$, for all $i$ such that $\Tr(\mathbf{V}_i^{(0)}\mathbf{V}_i^{(0)^H}) = P_{\max}$, we define layers $k = 1, ... K$ as,
\begin{equation}
\mathbf{U}^{(k)}_i \!=\! \bigg(\!\sum_{\substack{j}}\mathbf{H}_{ij}^{}\mathbf{V}_{j}^{(k-1)}\mathbf{V}_{j}^{(k-1)^H}\!\mathbf{H}_{ij}^{H} \!+ \!\sigma^2\mathbf{I}_R\!\bigg)^{-1}\!\!\!\!\mathbf{H}_{ii}^{}\mathbf{V}_{i}^{(k-1)}\!\! \label{E:unfold_u} 
\end{equation}
\begin{equation}
\hat{\mathbf{W}}^{(k)}_i \!=\! \big( \mathbf{I}_d - \mathbf{U}_i^{(k)^H}\mathbf{H}_{ii}^{}\mathbf{V}_{i}^{(k-1)} \big)^{-1} \label{E:unfold_w}
\end{equation}
\begin{equation}
 \hspace{4mm} \mathbf{\xi}^{(k)} = \Psi({\bbS},[\mathbfcal{U}^{(k)},\mathbfcal{V}^{(k-1)}]; \theta^{}) \hspace{-5mm} \label{E:unfold_xi} 
\end{equation}
\begin{equation}
{\mathbf{W}}^{(k)}_i = \Phi_{\mathbf{\xi}_i^{(k)}}(\hat{\mathbf{W}}^{(k)}_i) + \hat{\mathbf{W}}^{(k)}_i \label{E:learn_w}
\end{equation}
\begin{equation}
\bar{\mathbf{V}}^{(k)}_i \!=\! \bigg( \sum_{\substack{j}} \mathbf{H}_{ij}^{H}\mathbf{U}_j^{(k)}{\mathbf{W}}_j^{(k)}\mathbf{U}_j^{(k)^H}\mathbf{H}_{ij}^{} + \mu \mathbf{I}_T\bigg)^{-1}\!\!\! \mathbf{H}_{ii}^{H}\mathbf{U}_i^{(k)}{\mathbf{W}}_i^{(k)} \! \!\!\label{E:unfold_v}
\end{equation}
where updates~\eqref{E:unfold_u}-\eqref{E:unfold_v} are computed in parallel for every user $i$.

Here, $\Psi$ in~\eqref{E:unfold_xi} is a complex-valued GNN (CV-GNN) architecture with a set of trainable parameters $\theta$. 
The matrix {$\bbS \in \mathbb{C}^{M\times M}$} is obtained through a learnable transformation applied to the CSI tensor, enabling its use within the CV-GNN $\Psi$.  
This tensor transformation is to be described in more detail in~\eqref{eq:H_bar}. 
The trainable parameter $\mu \in \mathbb{C}$ resembles the Lagrange multiplier for the power constraint in the original WMMSE formulation. 
The output of each hybrid layer of our architecture is then given by $\bar{\mathbfcal{V}}^{(k)}$, such that $[\bar{\mathbfcal{V}}^{(k)}]_{i::} = \bar{\mathbf{V}}^{(k)}_i$ for all $i$. However, $\bar{\mathbfcal{V}}^{(k)}$ is the \emph{raw} transmitter-beamformer that does not necessarily obey the power constraint. 

\begin{figure*}
	\centering
	\includegraphics[width=0.80\linewidth]{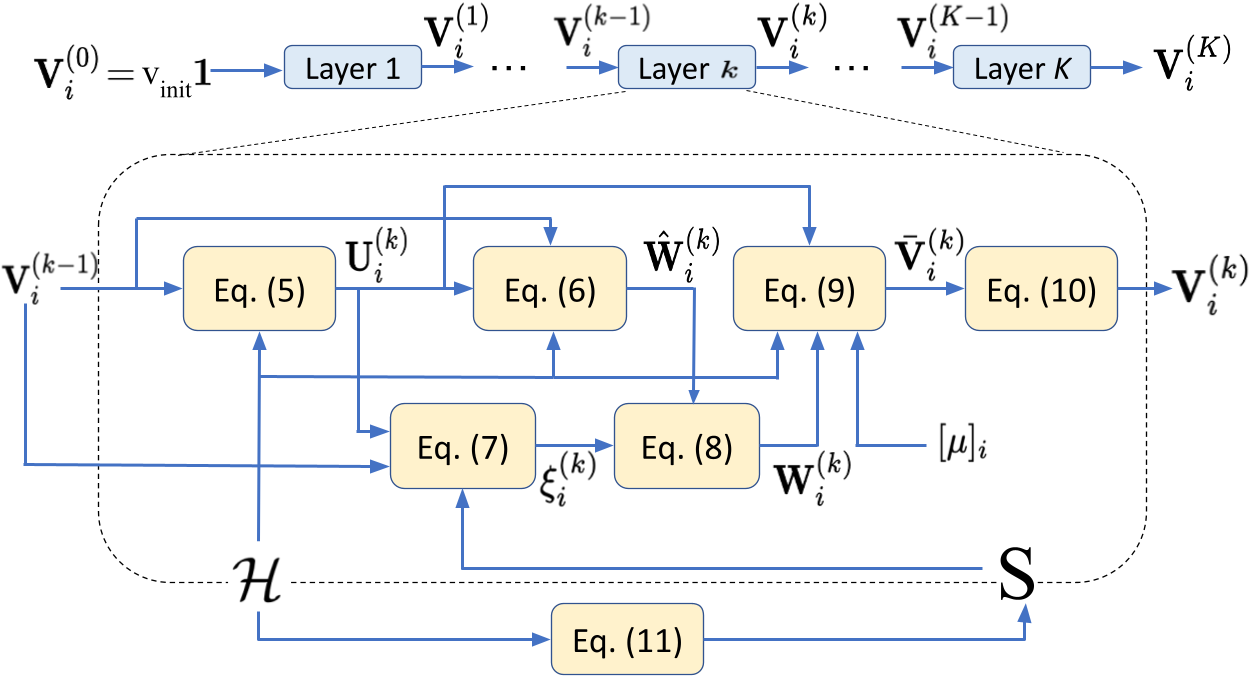}
	\vspace{-4mm}
	\caption{\small Flow diagram depicting the variable dependencies in any given intermediate layer $k$ of the proposed $K$-layered UWMMSE algorithm.
	Updates are shown for an arbitrary node $i$ and are computed in parallel for all $i$.
 The yellow blocks represent the five equations in \eqref{E:unfold_u}-\eqref{E:unfold_v} plus \eqref{eq:nonlinear_beta} and \eqref{eq:H_bar}.
	The input to the layered structure is $\bbV^{(0)}_i = \mathrm{v}^{}_{\mathrm{init}}\mathbf{1}_{T \times d} \,$ where $\mathrm{v}^{}_{\mathrm{init}} = \sqrt{\frac{P_{\max}}{2Td}}(1+\sqrt{-1})$ and the output transmitter-beamformer is given by $\bbV^{(K)}_i$ for all $i$. 
	Dependence on $\sigma$ and $\{
	\bbU_j,\bbW_j,\bbV_j\}$ for all $j\neq i$ is implicit.
}
	\vspace{-4mm}
	\label{F:unr}
\end{figure*}

To enforce the power constraint on the output of the feedforward architecture, we introduce a non-linear activation function $\beta (\cdot)$ in the hybrid layers of $\Lambda(\cdot;\Theta)$ which saturates the model output beyond the permissible values.    
For the multi-antenna setup that we consider here, this involves constraining $\mathbfcal{V}^{(k)}$ identically in each layer $k$ such that all its elements $\mathbf{V}_i^{(k)} = [\mathbfcal{V}^{(k)}]_{i::}$ satisfy $\Tr\left(\mathbf{V}_i^{(k)}\mathbf{V}_i^{{(k)}^H}\right) \leq P_{\mathrm{max}}$. 
To attain this, the activation $\beta$ in each layer $k$ for all $i$ is defined as
\begin{align}\label{eq:nonlinear_beta}
    \mathbf{V}_i^{(k)} &= \beta(\bar{\mathbf{V}}_i^{(k)}) \nonumber \\ 
    & = \begin{cases}
    \bar{\mathbf{V}}_i^{(k)},& \text{if } \Tr\left(\bar{\mathbf{V}}_i^{(k)}\bar{\mathbf{V}}_i^{{(k)}^H}\right) \leq P_{\mathrm{max}},\\
    \bar{\mathbf{V}}_i^{(k)} \cdot  \frac{\sqrt{ P_{\mathrm{max}}}}{\lVert\bar{\mathbf{V}}_i^{(k)}\rVert_F}, & \text{otherwise},
\end{cases}
\end{align}
where $||\cdot||_F$ denotes the Frobenius norm. We note that a non-linear mapping of this form was used in the PGD based beamforming strategy of~\cite{pellaco2020deep} as the projection step. A schematic view of the variable dependence of the proposed UWMMSE is given in Fig.~\ref{F:unr}.
%
%

It is essential to note here that if we ignore~\eqref{E:unfold_xi} and set $\Phi_{\mathbf{\xi}^{(k)}}(\cdot) = \mathbf{0}_{d\times d}$ for every layer $k$, then \eqref{E:unfold_u}-\eqref{eq:nonlinear_beta} boil down to the classical BCD closed-form updates of WMMSE.
However, by providing the additional flexibility to UWMMSE of learning a set of representations $\mathbf{\xi}^{(k)}$ in \eqref{E:unfold_xi} for each node -- which are implemented as parameters of a CV-MLP in~\eqref{E:learn_w} corresponding to the node -- we enable faster convergence and better performance compared with the classical WMMSE, as illustrated in Section~\ref{S:num_exp}.

In building~\eqref{E:unfold_u}-\eqref{E:unfold_v}, one of the primary design considerations is the choice of WMMSE variables that are to be learned. 
We choose to preserve the update structures of $\mathbfcal{U}$ and $\mathbfcal{V}$ since these are tightly related to the underlying communication dynamics of the wireless network.
Indeed, these two update equations explicitly quantify the effects of interference on transmitters and receivers.
On the other hand, $\hat{\mathbfcal{W}}$ is representative of the quality of the channel connecting a transmitter and its corresponding receiver and plays a key role in driving $\mathbfcal{V}$ and $\mathbfcal{U}$ to their near-optimal values. 
Our hypothesis is that if $\hat{\mathbfcal{W}}$ can be accelerated towards convergence through data-driven optimization, then it will lead $\mathbfcal{U}$ and $\mathbfcal{V}$ to faster convergence without affecting the dynamics of the wireless network.
Having thus finalized the variable to be augmented by learning, the next design consideration is the structure of the learnable transformation $\Phi$. 
To that end, we propose the use of a complex-valued multi-layer perceptron (CV-MLP) with a single hidden layer as $\Phi: \mathbb{C}^{d^2} \to \mathbb{C}^{d^2}$. 
On account of its universal approximation property (UAP), an MLP is capable of modelling any continuous and bounded function of arbitrary complexity~\cite{leshno1993multilayer}. 
Therefore, without imposing any additional inductive bias on the structure of the transformation, the proposed method provides the necessary capabilities for $\mathbfcal{W}$ to follow an improved update trajectory compared to that taken by $\hat{\mathbfcal{W}}$ alone. 

In any given layer $k$, parameters $\xi_i^{(k)}$ corresponding to the CV-MLP $\Phi_{\xi_i^{(k)}}$, defined on node $i$, are learned using a CV-GNN $\Psi$. 
In addition to the complex-valued parameters $\xi$, $\Phi$ also uses Cartesian non-linearites~\cite{kuroe2003activation} on both hidden and output layers, which are capable of handling complex-valued outputs. More specifically, it uses the ReLU family of activations~\cite{maas2013rectifier} that are applied independently on the real and imaginary components of the layer-wise outputs of $\Phi$, thereby transforming both magnitude and phase.
In essence, we propose a learnable transformation in each unfolded layer $k$ at two levels. Firstly, we leverage the UAP of the CV-MLP $\Phi$ to frame a general functional transformation for the receiver weights, and secondly, the node-specific parameters $\xi^{(k)}_i$ of the transformation $\Phi_{\xi^{(k)}_i}$ are learned using a CV-GNN $\Psi$ as representations for all nodes $i$.
While the standard practice is to learn the parameters of a CV-MLP directly, we take the aforementioned route as a standalone CV-MLP cannot generalize to arbitrary connectivity patterns in a wireless network graph. 
For instance, the learnable model must be capable of generating node representations by leveraging the local connectivity structure of the wireless network embedded in $\mathbf{S}$.
A generic GNN architecture, through a sequence of aggregation and transformation operations~\cite{kipf2016semi,gama2018convolutional}, is able to sufficiently capture this structure.
Moreover, GNNs are typically \emph{permutation equivariant}~\cite{kipf2016semi}, thus offering better generalization performance against variations in node ordering.
The choice of the specific GNN architecture and the size of the trainable parameters are, however, arbitrary and can be made depending on the nature of the problem. 
In this case, to ensure computational simplicity~\cite{dwivedi2023benchmarking}, we choose a complex-valued architecture inspired by a graph convolutional network~\cite{kipf2016semi} (CV-GCN) along with Cartesian non-linearites~\cite{kuroe2003activation} on $\Psi(\cdot;\theta)$, as described in~\eqref{E:psi}--\eqref{E:psi_2}. 

Independent of the choice of architecture $\Psi$, we treat the CSI tensor as a weighted adjacency structure of a directed graph which is used to aggregate information from the neighboring nodes~\cite{segarra_2017_optimal}. 
In the multi-antenna setting that we consider here, the link -- either the channel or interference -- between any transmitter $i$ and receiver $r(j)$ is described by an $R \times T$ matrix that depends on the number of transmitter and receiver antennas. 
However, since CV-GCN $\Psi$ requires the channel between $i$ and $r(j)$ to be represented by a scalar coefficient~\cite{kipf2016semi}, we propose the use of a single-layered $1\times 1$ depth-wise convolution~\cite{lin2013network} operation with shared filter parameters to transform $\mathbfcal{H}$ to an amenable structure. 
Essentially, we define an additional fully connected neural layer $\Gamma(\mathbfcal{H};\boldsymbol{\omega}): \mathbb{C}^{M\times M\times R \times T} \to \mathbb{C}^{M\times M\times 1}$, $\boldsymbol{\omega} \in \mathbb{C}^{R\times T}$ which generates a matrix $\bbS$ -- such that
\begin{equation}\label{eq:H_bar}
[\bbS]_{ij} = \sum_{p,q} \omega_{pq} H_{ijpq} \quad \text{for all} \,\, i,j
\end{equation}
-- which forms the input\footnote{In practice, we observed that a row-normalization operation on $\bbS$ prior to inputting it in~\eqref{E:unfold_xi} had the effect of stabilizing training and thereby improving the overall performance. See footnote 2 for access to implementation code.} to $\Psi(\cdot;\theta)$ in~\eqref{E:unfold_xi}. 
Indeed, this operation can be interpreted as a learnable weighted-combination of the $RT$ antenna coefficients for each channel matrix $\mathbf{H}_{ij} = [\mathbfcal{H}]_{ij::}$ to generate a scalar representation $S_{ij} = [\bbS]_{ij}$ for the channel. 

In addition to capturing the local connectivity structures in graphs, GNNs are also well suited to handle features or signals supported on the nodes of a graph \cite{kipf2016semi, marques_2016_sampling}.
More precisely, setting the aggregation matrix $\bbS \in \mathbb{C}^{M\times M}$ and defining a features matrix $\bbQ \in \mathbb{C}^{M \times F'}$, the GNN in~\eqref{E:unfold_xi} has the functional form $\Psi(\bbS, \bbQ; \theta)$ , where $\bbQ = [\mathbfcal{U}^{(k)},\mathbfcal{V}^{(k-1)}]$ is a concatenation of the current iterates for $\mathbfcal{U}$ and $\mathbfcal{V}$. 
Specifically, we consider the case where $d=1$ in the rest of the paper, resulting in $F'=R+T$. 
However, note that the model can be extended to the case where $d>1$ by a simple pooling transformation on the last dimensions of $\mathbfcal{U}$ and $\mathbfcal{V}$. 
Such a formulation explicitly couples the CV-GNN with the current state of the WMMSE variables.
This is essential to ensure that the CV-GNN output has a functional dependence on the optimization trajectory across layers.
Additionally, certain QoS metrics like traffic rates, user priority, queue lengths etc. can also constitute relevant node features in this case.
While the incorporation of the aforementioned or more node features in our model is a straightforward task, the detailed analyses -- theoretical and experimental -- of their effects on model performance is beyond the scope of this work.

Thus, having described the aggregator matrix $\bbS$ and the feature matrix $\bbQ$, we now present the exact architecture of $\Psi(\cdot;\theta)$ in the proposed model
\begin{align}
\Psi(\mathbf{S},\bbQ; \theta) & = \alpha_2 \left( \mathrm{diag}(\bbS) \, \bbZ \, \bbtheta_{21}^H + \bbS \, \bbZ \,\bbtheta_{22}^H \right), \label{E:psi}\\
\bbZ & = \alpha_1 \left( \mathrm{diag}(\bbS) \, \mathbf{Q} \, \bbtheta^H_{11} + \bbS \, \mathbf{Q} \,\bbtheta^H_{12} \right), \label{E:psi_2}
\end{align}
where $\theta = \{\bbtheta_{11}, \bbtheta_{12}, \bbtheta_{21}, \bbtheta_{22}\}$, and both $\alpha_1$ and $\alpha_2$ are Cartesian RELU activation functions.
Note here that we have an additional set of weights for the diagonal elements in the formulation of $\Psi(\cdot;\theta)$.
This is essentially to emphasize the importance of the transmission channel elements with respect to (w.r.t) the off-diagonal interference elements in the learnable model.

Finally, trainable parameter $\mu$ -- shared by all nodes -- occupies the place of the Lagrange multiplier in the original WMMSE formulation.
Its primary purpose is to incorporate the node-wise power constraint. 
Intuitively, a larger $\mu$ in~\eqref{E:unfold_v} would result in a $\bar{\bbV}$ such that $\Tr(\bar{\bbV}\bar{\bbV}^{H})$ is small even when the interference component is small. 
This ensures that $\bar{\bbV}$ is less likely to deviate far from the power constraint, even before it is enforced explicitly in~\eqref{eq:nonlinear_beta}. 
This allows the model to operate more frequently in the linear region of the activation $\beta$, leading to better numerically conditioned gradient propagation across layers. 
The parameter $\mu$ is trained directly using the gradient feedback from~\eqref{eq:loss_function}. 


\subsection{Permutation Equivariance of UWMMSE }\label{Ss:permEq}



Permutation equivariance is a property by virtue of which the performance of a GNN model remains consistent w.r.t variations in node identities.
It is of particular importance for dynamic WANETs wherein nodes may move in and out of the network or even operate with varying topologies. 
Moreover, the channel sensors may convey their estimates in different orders leading to a rearrangement of the rows and columns of $\bbS$.  
In all these cases, the GNN must be equipped to consistently maintain the quality of its predictions across the variations.
Since the fundamental learnable module $\Psi(\bbS,\bbQ;\theta)$ in our proposed model $\Lambda(\mathbfcal{H}; \Theta)$ -- where $\Theta = \{\theta,\boldsymbol{\omega},\mu\}$, such that $\boldsymbol{\omega}$ and $\mu$ are independent of node ordering -- is permutation equivariant~\cite{kipf2016semi}, it is essential to establish that this key property is inherited by the overall UWMMSE architecture.

Similar to~\cite{chowdhury2021unfolding, chowdhury2021efficient} we now formalize the definition of permutation equivariance. 
Let us consider a generic permutation matrix $\bbPi \in \{0,1\}^{M \times M}$. 
Further, let $\ccalF$ denote the set of all functions $f: \mathbb{C}^{M \times M} \to \mathbb{C}^M$.

\begin{mydefinition}
    A function $f \in \ccalF$ is permutation equivariant if $f(\bbPi \bbS \bbPi^\top) = \bbPi f(\bbS)$ for all permutations $\bbPi$ and all matrices $\bbS$.
\end{mydefinition}

Essentially, a certain permutation of the node indices of a graph, input to function $f$ which is permutation equivariant, applies the same permutation to the indices of the corresponding node outputs without altering their values.  
Next, we formally qualify the permutation equivariance of the proposed UWMMSE model conditioned on the specific $\Psi$ in~\eqref{E:unfold_xi}.

\begin{myproposition}\label{P:equiv}
The UWMMSE model $\Lambda(\cdot\,; \Theta)$ is permutation equivariant.
\end{myproposition}	

\begin{myproof}
    The proof is relegated to Appendix~\ref{app:permEq}.
\end{myproof}


\subsection{Training Process}\label{Ss:train}

Having explained the inner workings of the proposed UWMMSE for a given set of learnable parameters $\Theta = \{\theta,\omega,\mu\}$, we now shift focus to the training of the architecture.
Given a fixed $\Theta$, the model $\Lambda( \mathbfcal{H} ; \Theta)$ generates the transmitter-beamformer corresponding to CSI $\mathbfcal{H}$, which is used to obtain a network sum-rate utility given as $\sum_{i=1}^M c_i(\Lambda( \mathbfcal{H} ; \Theta), \mathbfcal{H})$; [cf.~\eqref{E:optimization_problem} for $\alpha_i=1$].
Thus, the loss can be defined as
\begin{equation}\label{eq:loss_function}
    \ell(\Theta) = - \mathbb{E}_{\mathbfcal{H} \sim \mathcal{D}} \left[ \sum_{i=1}^M c_i(\Lambda( \mathbfcal{H} ; \Theta), \mathbfcal{H}) \right],
\end{equation}
where $\mathcal{D}$ is the channel state distribution of interest. 
Even if $\mathcal{D}$ is known, minimizing $\ell(\Theta)$ w.r.t $\Theta = \{\theta,\omega,\mu\}$ is a non-convex problem.
However, notice that $\Psi(\cdot; \theta)$ in~\eqref{E:unfold_xi} and $\Gamma(\cdot; \omega)$ in~\eqref{eq:H_bar} are differentiable w.r.t $\theta$ and $\omega$, respectively.
Thus, given a set of $\mathbfcal{H}$ drawn from $\mathcal{D}$, we employ stochastic gradient descent to minimize~\eqref{eq:loss_function}.
Therefore, UWMMSE is essentially an unsupervised learning algorithm which only needs samples of the CSI $\mathbfcal{H}$ but does \emph{not} need the true-optimal transmitter-beamformers (labels) associated with those channels, which can be tremendously expensive to generate.
While another possibility is to use WMMSE power allocation as the training labels, this effectively limits the learning capacity of the proposed UWMMSE by the near-optimality of the WMMSE output.

\begin{remark}[Application to SISO systems]\label{R:comparison} 
\normalfont While our method is designed for beamforming in the more general MIMO setting, it can be seamlessly employed for power allocation in SISO wireless networks.
In the SISO setting, however, there would be no need for the neural network in~\eqref{eq:H_bar} since when $R=T=1$ the CSI matrix can directly be used as the generalized adjacency matrix in the GNN $\Psi$, thus reducing the trainable parameters only to $\Theta = \{ \theta, \mu\}$.
Similarly, intermediate variables $\mathbf{U}^{(k)}_i, \mathbf{W}^{(k)}_i, \mathbf{V}^{(k)}_i$ are reduced to scalars, but their update equations in~\eqref{E:unfold_u}-\eqref{E:unfold_v} remain valid and naturally present a lower computational load.
In this sense, for the SISO case, our proposed UWMMSE resembles the unfolding scheme presented in~\cite{chowdhury2021unfolding}.
However, in~\cite{chowdhury2021unfolding}, the embedded transformation $\Phi$ on $\mathbfcal{W}$ in each hybrid layer has a fixed affine structure that lacks the UAP of the CV-MLP.
Further, the authors in~\cite{chowdhury2021unfolding} consider only real-valued channel realizations which are constructed using the magnitude of the complex-valued channels.
Thus, their method completely ignores the phase information embedded in the channel coefficients thereby oversimplifying the problem. 
Also, each layer $k$ of the model presented in~\cite{chowdhury2021unfolding} has its own independent set of GNNs $\Psi(\cdot;\theta^{(k)})$ resulting in growing complexity of their proposed solution with increasing number of layers. 
More importantly, a model of this form requires the number of layers to be fixed at both training and inference making it inflexible at deployment.
In this respect, our more general UWMMSE framework for complex-valued MIMO still presents advantages w.r.t existing works even if we focus on the SISO setting.
\end{remark}

\subsection{UWMMSE convergence: A necessary condition}\label{Ss:convergence}

We theoretically establish the behavior of the UWMMSE architecture with an arbitrarily large number of layers.

\begin{mytheorem}\label{T:convergence_necessary}
    Consider a UWMMSE architecture~\eqref{E:unfold_u}-\eqref{eq:H_bar} of infinite depth -- where $\Phi_{\xi_{i}^{(k)}}(\cdot)$ are continuous functions for all $i,k$ -- being used for beamforming to transmit a signal $\mathbf{x}_i \in \mathbb{C}^1$ i.e., $d=1$. Consider the extreme low-noise regime so that $\sigma = 0$ and set $\mu=0$. Denote using $^*$ the optimal solutions to problem~\eqref{E:problem_reformulation}.
    If $\ \mathbf{V}_i^{{(k)}^H}\mathbf{V}_i^{(k)} \to P_i^*$ such that $0< P_i^* < P_{\max}\ $ for all $i$, uniformly as $k \to \infty$, then it must hold that
   \begin{align}\label{E:convergence_necessary_condition}
    &\bbA_i^{-1}\bigg[\sum_{\substack{j\neq i}}\mathbf{H}_{ij}^{H}\mathbf{U}_{j}^{(k)}(w_j^* \Phi_{\xi_{i}^{(k)}}(w_i^{(k)}) - w_i^* \Phi_{\xi_{j}^{(k)}}(w_j^{(k)})) \nonumber \\ &\quad\quad\quad\quad{\mathbf{U}_{j}^{(k)}}^H\!\mathbf{H}_{ij}^{}\bigg]\bbC_i^{-1}\bar{\bbB}_i \to \bb0  \,\,\,\, \text{for all} \,\, i \,\, \text{as} \,\, k \to \infty
    \end{align}
    where $\bbW \in \mathbb{C}^{1\times1}$ is represented as the scalar $w \in \mathbb{C}$, $\bbA_i =  \!\sum_{\substack{j}}\mathbf{H}_{ij}^{H}\mathbf{U}_{j}^{(k)}w_{j}^{(k)}{\mathbf{U}_{j}^{(k)}}^H\!\mathbf{H}_{ij}^{}$, $\bar{\bbB}_i = \mathbf{H}_{ii}^{H}{\mathbf{U}_{i}^{*}}$, and $\bbC = \sum_{\substack{j}}\mathbf{H}_{ij}^{H}\mathbf{U}_{j}^{*}w_j^{*}{\mathbf{U}_{j}^{*}}^H\!\mathbf{H}_{ij}^{}$.
\end{mytheorem}

\begin{myproof}
    The proof is relegated to Appendix~\ref{app:thrm}.
\end{myproof}

Theorem~\ref{T:convergence_necessary} states that if UWMMSE learns the optimal transmitter beamformer $\mathbf{V}_i^*$ -- where, $\bbV_i^{*^{H}}\bbV_i^* = P_i^*$ -- uniformly at deeper layers, then the learned transformation $\Phi_{\xi_i}(\cdot)$ must satisfy~\eqref{E:convergence_necessary_condition} asymptotically for all $i$.
Notice that  $ \Phi_{\xi_i^{(k)}}(w_i^{(k)}) \to 0$ or, more generally, $ \Phi_{\xi_i^{(k)}}(w_i^{(k)}) \to \delta w_i^{(k)}$ for some constant $\delta$ satisfy the requirement in~\eqref{E:convergence_necessary_condition}.
This is intuitively pleasing, since this limiting behavior of $ \Phi_{\xi_i^{(k)}}(w_i^{(k)})$ implies that deeper layers of UWMMSE would resemble the classical iteration of WMMSE, for which we know that the optimal beamformer is a fixed point~\cite{shi2011iteratively}.
In other words, the proposed learnable module is sufficiently expressive to modify the updates for the first few layers to accelerate convergence while recovering the optimal asymptotic guarantees of the classical WMMSE algorithm at deeper unfolded layers.
Finally, note that Theorem~\ref{T:convergence_necessary} is independent of the choice of the specific CV-GNN $\Psi(\cdot;\theta)$ in~\eqref{E:unfold_xi}. 
Therefore, we can safely claim that the aforementioned result is an attribute of the hybrid model in~\eqref{E:unfold_u}-\eqref{eq:nonlinear_beta},
and is independent of specific the CV-GNN architecture chosen to learn the parameters $\xi$. 


\subsection{Parameter sharing for computational efficiency}\label{Ss:parametersharing}

\normalfont The CV-GNN $\Psi(\cdot;\theta)$ is shared by all unfolded layers, i.e., $\theta$ does not depend on the layer index $k$ in~\eqref{E:unfold_xi}. 
Such a scheme ensures that $\theta$ is trained using gradient feedback that accumulates across layers and depends on the overall optimization trajectory. 
Moreover, a formulation of this form allows for flexibility in adding or removing unfolded layers at deployment (after training has been completed). 
Another immediate advantage is an $\mathcal{O}(K)$ reduction in the number of trainable parameters with respect to the layer-dependent alternative, making the training process less computationally expensive and time-consuming.
Further, the trainable parameters $\omega$ of the tensor-transformation $\Gamma(\cdot;\omega)$ in~\eqref{eq:H_bar} are identical for all channel elements. 
This is appropriate as all channel representations must have identical functional mapping from their respective antenna coefficients in a way that is analogous to shared $1\times 1$ convolutions of image pixels. 
Additionally, having a shared filter kernel allows for an $\mathcal{O}(M^2)$ reduction in the number of trainable parameters.
Finally, note that $\mu$ is also tied across layers, further reducing the number of trainable parameters. 


\subsection{Complexity analysis, scalability and distributed implementation}\label{Ss:distributed}

Per iteration computational complexity of WMMSE~\cite{shi2011iteratively} is $\mathcal{O}(M^2TR^2 + M^2RT^2 + M^2T^3 + M^2R^3)$ for an $M$-user interference network with $R$ receiver antennas and $T$ transmitter antennas.
It can be re-written as $\mathcal{O}(M^2[\max\{R,T\}]^3)$.
Each unfolded layer in UWMMSE inherits this complexity directly as they perform the same update as WMMSE. 
Moreover, the trained CV-GCN in each unfolded layer incurs a feedforward complexity of $\mathcal{O}(M^2F)$ where the hidden layer size is $F$ \cite{kipf2016semi}.
Finally, each trained single-hidden-layered CV-MLP has a feedforward complexity of $\mathcal{O}(MG)$, where the hidden layer size is $G$.  
Hence, the total complexity of the trained UWMMSE is given as $\mathcal{O}(M^2[[\max\{R,T\}]^3 + F] + MG)$. 
Clearly, for a fixed set of antenna sizes $\{R,T\}$ and pre-designed hidden dimensions $F$ 
 and $G$, per layer complexity of UWMMSE varies as $\mathcal{O}(M^2)$ w.r.t network size $M$. 
 This is same as the per-iteration complexity of WMMSE with fixed antenna size. 
Consequently, by truncating the number of unfolded layers $K$ in UWMMSE as compared to the number of iterations in WMMSE, the inference time is significantly reduced.  
This will be empirically validated in Section~\ref{Ss:performance_comparison}. 

In addition to the computational complexity, it is also important to note the size of the model given by the number of its trainable parameters. 
The proposed architecture has very few trainable parameters making it easy to train, and likely to generalize as illustrated in Section~\ref{S:num_exp}. 
The number of parameters $\theta$ in the $2$-layered CV-GCN $\Psi$ is $\mathcal{O}(F[F'+G])$, where $F'$ is the input size.  
Further, the linear layer $\Gamma$ has $\mathcal{O}(RT)$ trainable parameters and $\mu$ has a size of just $1$. 
Thus, the total number of trainable parameters of UWMMSE is $\mathcal{O}(RT + F[F'+G])$, and is independent of the number of users $M$.
As a result, the same model can be employed to process wireless networks of varying size with the assumption that the underlying channel model is identical.

While it is necessary to train the proposed UWMMSE in a centralized manner under the assumption that the centralized trainer has access to the full CSI tensor, the trained UWMMSE can support a distributed deployment  with only local information available at each node. 
This is mainly possible given the fact that power allocation at a given node $i$ depends only on the row slice $\mathbfcal{H}_{r(i):::}$ and the column slice $\mathbfcal{H}_{:i::}$ of the CSI tensor. 
Nevertheless, to achieve a fully distributed deployment, three vital assumptions are necessary, two of which are inherited directly from the distributed version of WMMSE~\cite{shi2011iteratively}. 
Firstly, for all receivers $r(j)$ where $j=1\dots M$, the local channel state estimates $\mathbf{H}_{ji}$ should be available to each transmitter $i$. 
Secondly, a mechanism is required to facilitate information feedback from receivers to all transmitters. 
These assumptions essentially enable the transmitters to compute $\mathbf{V}_i^{(k)}$ after receiving the corresponding $\mathbf{U}_i^{(k)}$ and $\mathbf{W}_i^{(k)}$ from each receiver $r(i)$ in all unfolded layers $k$. 
Finally, a copy of the full set of trained parameters $\Theta = \{\theta,\omega,\mu\}$ must be available to each node. 
Note that, while we can achieve distributed deployment in this manner, the feedback links will add to the communication overhead and therefore the inference time would be higher than that of the centralized version. 


\section{Numerical Experiments}\label{S:num_exp}
 
In this section, we present comprehensive numerical experiments to demonstrate the performance of the proposed UWMMSE model in allocating power to complex-valued MU-MIMO WANETs operating under various fading conditions and topologies.\footnote{We have released our code for this work at \href{https://github.com/ArCho48/Unrolled-WMMSE-for-MU-MIMO}{https://github.com/ArCho48/Unrolled-WMMSE-for-MU-MIMO}.}
A detailed description of the datasets is provided in Section~\ref{Ss:data} while the model architecture, hyper-parameters and system setup are presented in Section~\ref{Ss:model}. 
In Section~\ref{Ss:performance_comparison}, we compare our model performance with WMMSE and its truncated version, in terms of achieved sum-rate and allocation time.
In Section~\ref{Ss:net_gen} and Section~\ref{Ss:dist_gen}, we evaluate the generalization performance of our model across different operating conditions in training and inference.
Further, in Section~\ref{Ss:conv}, we present an illustration of the convergence behavior of our model.
Finally, in Section~\ref{Ss:robust} we investigate the robustness of our proposed model against norm-bounded distortions in the input CSI tensor. 

\subsection{Datasets}\label{Ss:data}

We use randomly generated \emph{geometric} channel realizations to evaluate the model performance. 
A geometric channel model has a composite structure with path loss and fading components. 
To simulate that, we construct a 2-D geometric graph with $M$ randomly sampled transceiver pairs. 
All transmitters and receivers are dropped uniformly at random at location $\mathbf{t}_i \in [0, \sqrt{M}]^2$ and $\mathbf{r}_i \in [0, \sqrt{M}]^2$. 
Path gain between transmitter $i$ and receiver $r(j)$ is then computed as an inverse function of their corresponding physical distance $l_{ij}$. We set the number of antennas as $R=3$ and $T=5$ for all the experiments.
For simplicity, we assume that a scalar complex-valued signal is being transmitted, i.e., $d=1$. 

Similar to~\cite{sun2021learning,sun2018learning,liang2019towards,chowdhury2021ml}, we choose the following fading channel models:

\noindent \textbf{Rayleigh}: For each channel matrix $\mathbf{H}_{ij}$ corresponding to the transceiver pair $ij$, we generate Rayleigh channel coefficients $[\mathbf{H}_{ij}]_{rt}$ independently for all antenna pairs $(r,t)$ as the real and imaginary components sampled independently from a standard normal distribution. 
Incorporating the path loss component, elements of the channel matrix $[\mathbf{H}_{ij}]_{rt}$ are given by
\begin{align*}
    [\mathbf{H}_{ij}]_{rt} &= \frac{1}{\sqrt{2}(1 + l_{ij}^3)}(a + \sqrt{-1}b) \quad \forall\  r,t.\\
    &\text{where, } a \sim \mathcal{N}(0,1), b \sim \mathcal{N}(0,1)
\end{align*}

\noindent \textbf{Rician}: For each channel matrix $\mathbf{H}_{ij}$ corresponding to the transceiver pair $ij$, we generate Rician channel coefficients $[\mathbf{H}_{ij}]_{rt}$ with $20$ dB $\mathcal{K}$-\textit{factor}{~\cite{tse2005fundamentals}} independently for all antenna pairs $(r,t)$ as the real and imaginary components sampled independently from a normal distribution. 
Incorporating the path loss component, elements of the channel matrix $[\mathbf{H}_{ij}]_{rt}$ are given by
\begin{align*}
    [\mathbf{H}_{ij}]_{rt} &= \frac{1}{(1 + l_{ij}^3)}(a + \sqrt{-1}b) \quad \forall\  r,t.\\
    &\text{where, } a \sim \mathcal{N}(\mu_{ric},\sigma_{ric}), b \sim \mathcal{N}(\mu_{ric},\sigma_{ric})
\end{align*}
where $\mu_{ric} = \sqrt{\frac{k}{2(k+1)}}$ and $\sigma_{ric} = \sqrt{\frac{1}{2(k+1)}}$ with $k=100$. 

\subsection{Model architecture}\label{Ss:model}
Our proposed feedforward UWMMSE architecture is composed of $3$ unfolded-WMMSE layers with a $2$-layered CV-GCN -- shared by the unfolded layers -- modeling the function $\Psi$ in~\eqref{E:unfold_xi}.
We set the hidden layer of CV-GNN as $F=32$ and that of CV-MLP as $G=16$.
The model consists of $3302$ trainable parameters.
NovoGrad~\cite{ginsburg2019stochastic} optimizer is employed for training across a maximum of $15000$ iterations on a batch of $64$ randomly sampled channel realizations with early stopping. 
The initial learning rate is set to \num{1e-2}. 
An interesting observation was that while $3$ unfolded-layers offered the best performance trade-off in terms of sum-rate and time at inference, the best training performance was achieved with just $1$ unfolded layer.
Recalling that our model is flexible insofar as to have different numbers of unfolded-layers at training and inference on account of parameter-sharing, we established a model setup wherein UWMMSE is trained with a single layer and then $2$ more layers are appended at inference, which share the learned components. 
Experimental results presented in this section demonstrate the effectiveness of this setup.  
At inference, we average the achieved sum-rate over $10000$ channel realizations for all experiments.
Our model, on account of being lightweight, is perfectly suited for both CPU and GPU operating environments.
Nevertheless, for uniformity and reproducibility, all experimental results for this paper are generated on an Nvidia GeForce RTX 2080 GPU. 

\subsection{Performance Comparison}\label{Ss:performance_comparison}

\begin{figure*}
	\centering
	\begin{subfigure}[]{
			\centering
			\includegraphics[width=0.30\textwidth, height=0.25\textwidth]{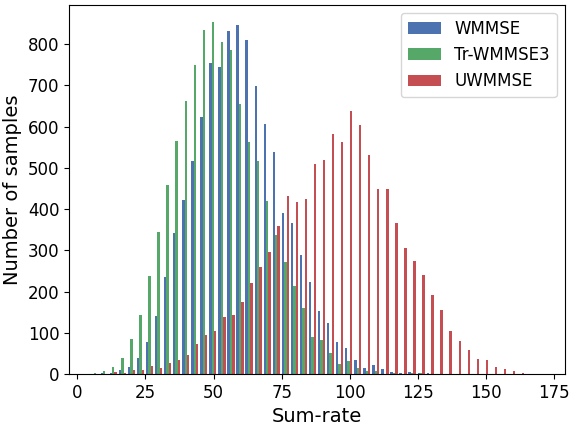}
			\label{Fig:performance_comparison_hist}}
	\end{subfigure}
	\begin{subfigure}[]{
			\centering
			\includegraphics[width=0.30\textwidth, height=0.25\textwidth]{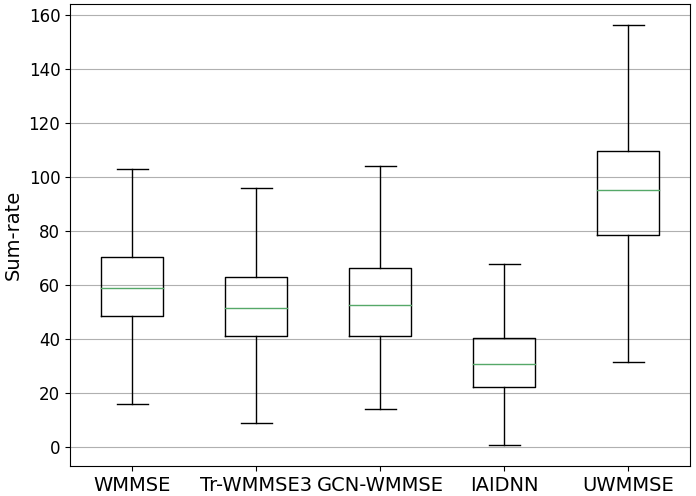}
			\label{Fig:performance_comparison_rayleigh}}
	\end{subfigure}	
	\begin{subfigure}[]{
			\centering
			\includegraphics[width=0.30\textwidth, height=0.25\textwidth]{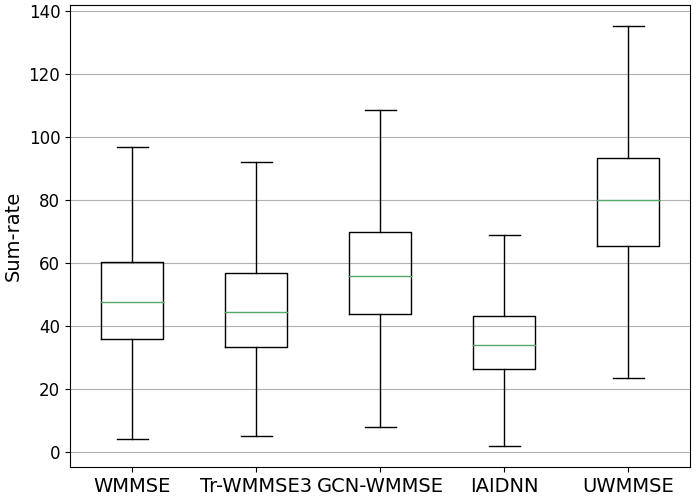}
			\label{Fig:performance_comparison_rician}}
	\end{subfigure}
		\vspace{-3mm}
		\caption{\small{Comparison of the achieved sum-rate by the proposed UWMMSE with full WMMSE \cite{shi2011iteratively}, a truncated version of it, IAIDNN \cite{hu2020iterative}, and GCN-WMMSE\cite{schynol2022coordinated}. 
		(a)~Sum-rate histogram for $\sim 10000$ Rayleigh CSI samples in the low-noise regime ($\sigma =  \num{2.6e-5}$). UWMMSE achieves significantly better performance than WMMSE while requiring equal number of iterations as the truncated version of it. 
		(b)~Box plots corresponding to histograms in (a), with additional comparisons with IAIDNN and GCN-WMMSE on identical CSIs.
		(c)~Counterpart of (b) for Rician channels.}
		}
		\vspace{-4mm}
		\label{Fig:performance_comparison}
\end{figure*}

The sum-rate performance of the proposed UWMMSE is compared with that of existing baselines including classical WMMSE and state-of-the-art connectionist methods.
For these comparisons, we had to choose an operating point between two regimes based on additive channel noise.
Firstly, in the \emph{low-noise} regime, the channel noise power is set at $-90 \dB$ or less. This is a more challenging scenario since the effects of interference tend to dominate that of channel noise and therefore the sum-capacity achieved depends largely on the precise beamforming at each transmitter. 
On the other hand, the \emph{high-noise} regime (noise power is set at $0 \dB$), offers a simplified setting wherein the achievable sum-capacity is innately low on account of high noise and therefore the exact beamforming at the transmitters is not critically important. 
In fact, our observation is that in this regime, the transmitters often exhibit binary characteristics, either transmitting at full power or not transmitting at all.
For our experiments, we choose to operate in the low-noise regime, specifically at $\sigma = \num{2.6e-5}\ (-114 \dB)$ since it allows us to evaluate the full potential of the proposed model. 

We now present a list of methods that we choose for performance comparison. All these methods address the common problem of beamforming in CV MU-MIMO WANETs.

\begin{enumerate}
    \item \textit{WMMSE} \cite{shi2011iteratively} is the classical baseline for our experiments as our method is an unfolded extension of it and is potentially an improvement over it. The maximum iterations per sample for WMMSE is set to $100$.
    \item \textit{Truncated WMMSE} (Tr-WMMSE) offers an empirical lower bound for UWMMSE in terms of the performance that can be achieved by equal number of WMMSE iterations as the number of unfolded layers without any learning. Tr-WMMSE is allowed $3$ iterations to match the number of UWMMSE layers.
    \item \textit{IAIDNN}~\cite{hu2020iterative} is a deep-unfolding framework to solve the sum-rate maximization problem for precoding design in MU-MIMO systems.
    \item \textit{GCN-WMMSE}~\cite{schynol2022coordinated} is a graph based unfolding framework for transceiver design in multicell MU-MIMO interference channels with local channel state information.  
\end{enumerate}

It is important to note here that both IAIDNN~\cite{hu2020iterative} and GCN-WMMSE~\cite{schynol2022coordinated} are unfolded extensions\footnote{For experimenting with IAIDNN and GCN-WMMSE, we have used the implementations by \textit{Schynol et al.} found at \href{https://github.com/lsky96/gcnwmmse.git}{https://github.com/lsky96/gcnwmmse.git}.} of WMMSE and therefore belong to a very specific class of hybrid algorithms that the proposed UWMMSE is also a part of. The main difference of UWMMSE w.r.t these methods lies in the structure of the respective unfolded components and the choice of the corresponding learnable modules (see Section~\ref{S:uwmmse}). 

The comparisons are shown in Fig~\ref{Fig:performance_comparison}. 
At any given instant, the channel conditions are randomly sampled from a fading distribution.
As a result, the sum-rate utility, which is conditioned on CSI, can vary significantly based on the exact samples used for evaluating different beamforming algorithms. 
A particular instance of CSI can be easy or hard to solve depending on the interference conditions and how these conditions contrast with the channel intensities.
However, the general performance over a large set of test samples should reveal the superiority of a particular beamforming algorithm over others.
This is illustrated in form of a histogram of achieved sum-rate by the full test set in a Rayleigh channel setting; see Fig~\ref{Fig:performance_comparison_hist}. 
The observed empirical distribution of sum-rate values over multiple channel realizations clearly reveals that UWMMSE significantly outperforms WMMSE for most realizations, with only a small overlap. 
Gain in full WMMSE performance compared to Tr-WMMSE, while expected due to difference in number of iterations, is not significant.
Intuitively, this is perhaps an indication that the WMMSE iterates approach a local optimum of the sum-rate objective fairly quickly and then simply converge to it over the rest of the gradient steps.
On the other hand, the UWMMSE iterates, supported by the embedded learnable components, take data-driven modified gradient steps to converge to a better local optimum within fewer steps.

Having established the superiority of UWMMSE w.r.t WMMSE, we now extend our investigation to the class of hybrid algorithms specifically under a Rayleigh channel setting; see Fig~\ref{Fig:performance_comparison_rayleigh}.
IAIDNN~\cite{hu2020iterative}, which uses learnable parameters to approximate matrix multiplication and inversion steps in WMMSE update rules, falls short of WMMSE by a significant margin with an average sum-rate of $32.17$. 
We suspect that the connectionist components of IAIDNN fail to leverage the inherent graph structure of the wireless networks and while its current formulation works well for fading channels without the path loss component (essentially lacking the geometric structure) and under a simplified high-noise setting, it is not equipped to deal with the full geometric channel model in a challenging low-noise scenario.
This shortcoming is addressed by the more recent GCN-WMMSE~\cite{schynol2022coordinated}, that uses a combination of graph filters and GCN~\cite{kipf2016semi} based graph learners to approximate certain WMMSE variables. 
Clearly, the use of graph information provides a big boost in its performance (average sum-rate of $55.19$) w.r.t IAIDNN.
Nevertheless, it also falls short of WMMSE, albeit marginally, in this particular setting. 
Clearly, the proposed UWMMSE is the superior algorithm among all the compared approaches. This superiority can be largely attributed to the two-step learning scheme wherein the first step leverages the underlying graph structure in the geometric wireless network to learn a set of parameters and the second step enforces a general functional transformation on a key WMMSE variable based on these learnt parameters (ref. Section~\ref{Ss:unfold}), thus facilitating a better convergence than the other approaches. 

We now shift focus to the Rician channel setting. 
Our objective in choosing the two different channel models for comparison is to demonstrate that the superiority of our method and the general performance trend for the various beamforming algorithms are not dependent on a particular channel type; see Fig~\ref{Fig:performance_comparison_rician}. 
While we observe a similar performance trend wherein the proposed UWMMSE beats all other methods comfortably, GCN-WMMSE~\cite{schynol2022coordinated} surpasses WMMSE marginally in this case and also closes the gap slightly with UWMMSE. 
IAIDNN~\cite{hu2020iterative} still lags behind all the methods but does marginally better than its Rayleigh counterpart.

\begin{table}[!htbp]
\centering
\caption{Time comparisons of all methods}
\begin{tabular}{|l | c | c | c | c | c|}
\hline
Algorithm  & Training  & Test \\
& time (min) & time (sec) \\
\hline \hline
WMMSE~\cite{shi2011iteratively} &  -  &  $1.305$ \\ 
Tr-WMMSE &  -  &  ${0.047}$ \\
IAIDNN~\cite{sun2018learning} & $\sim {10}$ & $0.64$ \\
GCN-WMMSE~\cite{hu2020iterative} &  $\sim {21}$  &  $1.365$ \\
UWMMSE & $\sim {35}$ & $0.054$ \\
\hline
\end{tabular}
\label{tab:performance}
\end{table}

While achieving a high sum-rate is the primary objective of the hybrid beamforming algorithms, it is also important for these methods to offer rapid inference as the beamforming process has to be capable of operating on the same time scale as potentially quickly varying channels. 
We therefore consider it essential to compare the time complexity of the aforementioned algorithms to generate the beamforming output for any given CSI input. 
A comparison of computation time is provided in Table~\ref{tab:performance}.  
Per sample inference time of UWMMSE is $54$ ms, which is $24$X lower than that of WMMSE which clocks around $1.30$ sec. 
Inference time of UWMMSE is predictably similar to that of Tr-WMMSE ($47$ ms per sample) since they have the same number of iterations, however, the learnable components of UWMMSE add slightly to its time complexity.
IAIDNN~\cite{hu2020iterative} has an inference time that is an order-of-magnitude higher than UWMMSE whereas GCN-WMMSE~\cite{schynol2022coordinated} takes the longest ($1.36$ sec per sample) inference time among all the methods.
Moreover, neither IAIDNN nor GCN-WMMSE achieve the same average performance as the proposed UWMMSE in their respective time duration.
Further, all the hybrid methods have a training component -- unlike the classical approach -- which tends to be time-consuming [ref. Table~\ref{tab:performance}].
Clearly, the training for IAIDNN is the fastest while the proposed UWMMSE takes the longest to train among all the hybrid methods. 
However, since training is one-time and is typically done prior to deployment, this is generally not a major concern for most applications.
We observe similar trends in both Rayleigh and Rician fading cases but we only present the Rayleigh case in Table~\ref{tab:performance}, for the sake of brevity. 

\subsection{Generalization across network sizes and fading types}\label{Ss:net_gen}

\begin{figure*}
	\centering
	\begin{subfigure}[]{
			\centering
			\includegraphics[width=0.30\textwidth,height=0.25\textwidth]{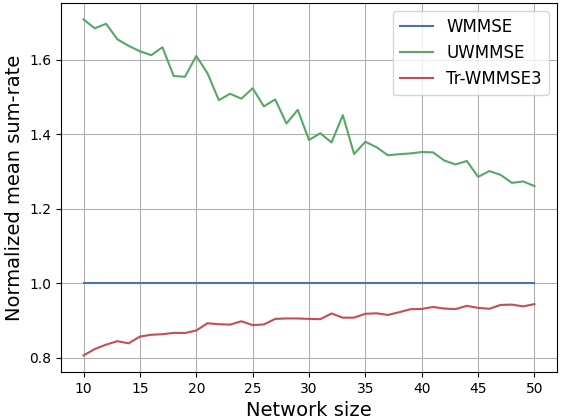}
			\label{Fig:generalization_intp}}
	\end{subfigure}
	\begin{subfigure}[]{
			\centering
			\includegraphics[width=0.30\textwidth, height=0.25\textwidth]{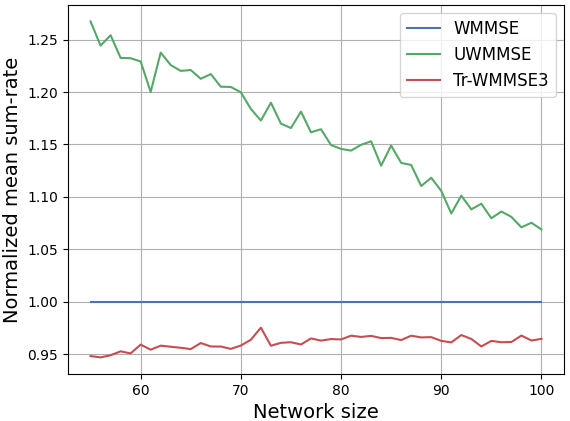}
			\label{Fig:generalization_exp}}
	\end{subfigure}	
	\begin{subfigure}[]{
			\centering
			\includegraphics[width=0.30\textwidth, height=0.25\textwidth]{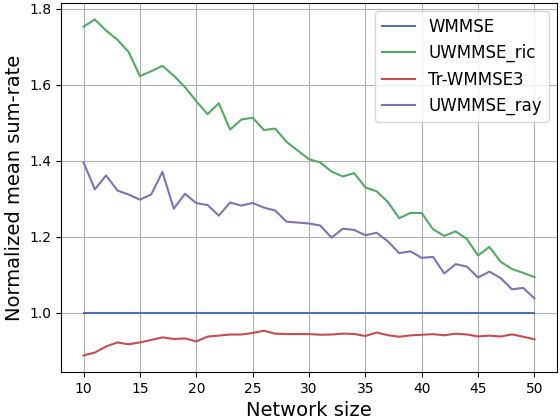}
			\label{Fig:generalization_ray_ric}}
	\end{subfigure}
		\vspace{-3mm}
		\caption{\small {Generalization performance of UWMMSE -- normalized w.r.t the corresponding WMMSE performance -- across multiple network sizes and fading models. 
		(a)~Normalized mean sum-rate achieved by UWMMSE on Rayleigh channel realizations with test sizes varying in the range $\{10, 11, \ldots, 50\}$, while training sizes were in the range $\{10, 12, 14, \ldots, 50\}$. 
        The normalized mean sum-rate achieved by Tr-WMMSE for the same range of network sizes is shown for comparison. 
		(b)~Counterpart of (a) but for test sizes varying in the range $\{55, \dots, 100\}$, with training sizes still in the range $\{10, 12, 14, \ldots, 50\}$. 
		(c)~Counterpart of (a) but for Rician channel realizations with an additional plot of normalized sum-rate achieved by a UWMMSE model trained on Rayleigh channel realizations and tested on Rician channel realizations. 
		}
		}
		\vspace{-4mm}
		\label{Fig:generalization}
\end{figure*}

Wireless networks are typically dynamic in terms of size, fading conditions, and channel noise power, among multiple other aspects that evolve through time.
Thus, we are interested in models that work for multiple network conditions.
To quantify this aspect, we evaluate the model on its generalization performance under a set of operating conditions at inference that are different from those at training.
Specifically, we perform this evaluation w.r.t network size and fading channel type as shown in Fig~\ref{Fig:generalization}.
We choose network size as one of the dynamic quantities since it is very common for a wireless ad-hoc network to have new nodes added to it or existing nodes removed from it.
The choice of fading channel type as the second dynamic setting is motivated by the fact that an ad-hoc network can group and re-group under various geographical (rural, urban, suburban) and climatic conditions such that there may or may not be dominant line-of-sight paths among transmitters and corresponding receivers.

Firstly, as a means of evaluating the interpolation behavior of UWMMSE across network sizes, we train it on even-valued sizes between $\{10,\ldots,50\}$ and then test it on all sizes between $10$ and $50$ for a Rayleigh channel setting. 
As shown in Fig~\ref{Fig:generalization_intp}, the model performs significantly better -- more than $1.2$-\textit{times} the corresponding WMMSE sum-rate -- for all sizes including the odd-valued sizes that were not available during training.
The observed drop in the model performance with increase in network size is expected since a larger network offers more interference at each transceiver and essentially poses a more challenging problem for the beamforming algorithm. 
Next, to evaluate the extrapolation behavior of UWMMSE, we test an identically trained model from the previous experiment on all sizes between $55$ and $100$.
As shown in Fig~\ref{Fig:generalization_exp}, the model performs reasonably well -- more than $1.05$-\textit{times} the corresponding WMMSE sum-rate -- for all unseen sizes.
Similar to the interpolation setting, a steady decrease in mean performance is observed in the extrapolation setting mainly due to increase in network size and the departure from the training setting.
However, the fact that the UWMMSE performance for all sizes between $10$ and $100$ is strictly above WMMSE irrespective of training conditions demonstrates the generalizability of our model to variations in network size.

Fig~\ref{Fig:generalization_ray_ric} illustrates the generalization behavior of UWMMSE to variation in fading conditions. 
A UWWMSE model trained on Rayleigh setting (UWMMSE\_ray) -- identical to the interpolation experiment -- is employed for inference on a Rician setting.
While UWMMSE\_ray fails to beat UWMMSE\_ric (which is specifically trained on a Rician channel setting), it still manages to comfortably beat the corresponding WMMSE performance for all network sizes.
Clearly, the proposed UWMMSE can reasonably generalize to different fading channel settings -- when trained under a fixed fading condition -- without any re-training on other fading conditions.


\subsection{Generalization across spatial distributions}\label{Ss:dist_gen}

So far, we have considered a WANET setting, wherein the transceivers are dropped uniformly at random in a square region of area $M$ as discussed in Section~\ref{Ss:data}. 
This choice of distribution is arbitrary and was made for the sake of simplicity.
Different real-world situations may yield other spatial distributions depending on topological conditions or mission-specific requirements.
For example, an Army unit might be deployed in a manner such that there are more soldiers stationed at a particular point-of-interest and their concentration reduces with distance from that point.
Such a deployment can be modelled more appropriately as a Gaussian distribution.
It is important that the proposed UWMMSE is able to handle such CSI tensors at inference, even when trained under a different distribution.
To that end, we take a UWMMSE model trained on uniformly distributed transceivers and evaluate its generalization performance on Gaussian distributed transceivers with a controlled standard deviation parameter.
As shown in Figure~\ref{Fig:dist_gen}, the performance of UWMMSE on the Gaussian test set is equivalent to the uniform test set only when the standard deviation is large enough at which point the Gaussian distribution is wide enough over the square region to essentially emulate a uniform distribution.
There is a clear degradation in performance with decreasing standard deviations as the nodes come closer together generating stronger interference.
Nevertheless, it is important to note that the mean sum-rates, normalized w.r.t the corresponding WMMSE sum-rates, are always above $1.0$. Hence, although UWMMSE achieves best possible performance when it sees identical distributions in training and inference, it generalizes reasonably well as compared to WMMSE even for unseen distributions at inference.

\subsection{Convergence}\label{Ss:conv}

\begin{figure*}
	\centering
 	\begin{subfigure}[]{
			\centering
			\includegraphics[width=0.30\textwidth, height=0.25\textwidth]{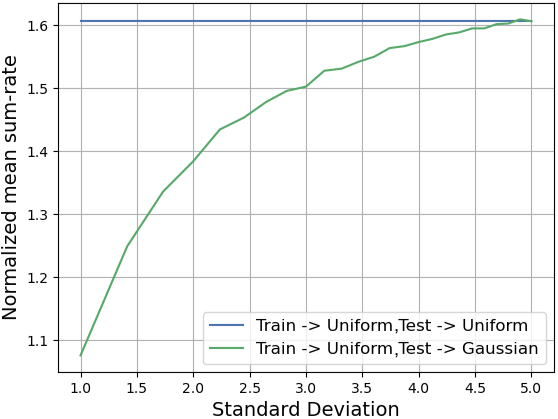}
			\label{Fig:dist_gen}}
	\end{subfigure}
	\begin{subfigure}[]{
			\centering
			\includegraphics[width=0.30\textwidth,height=0.25\textwidth]{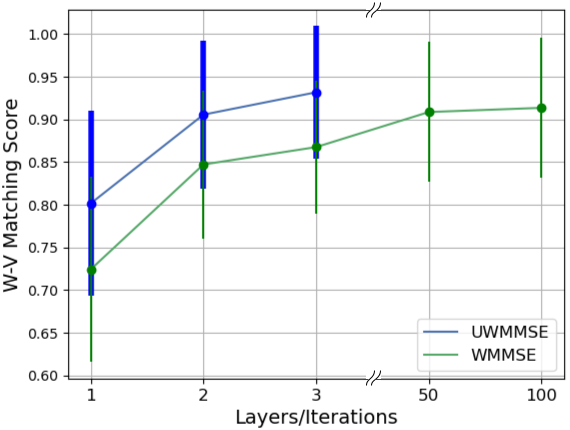}
			\label{Fig:convergence}}
	\end{subfigure}
	\begin{subfigure}[]{
			\centering
			\includegraphics[width=0.30\textwidth, height=0.25\textwidth]{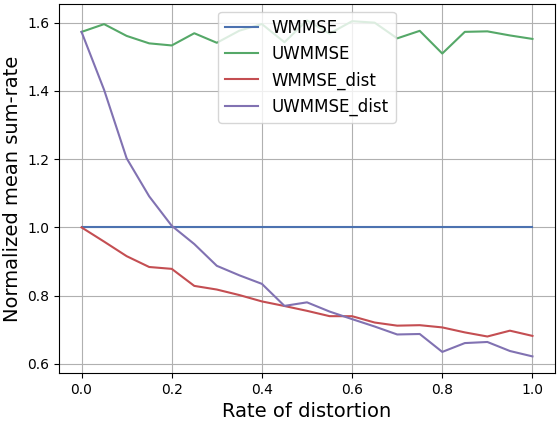}
			\label{Fig:robustness}}
	\end{subfigure}	
		\vspace{-3mm}
		\caption{\small {Performance generalization across spatial distributions, convergence, and robustness results.
        (a)~Variation in normalized sum-rate utility achieved by the UWMMSE model trained on uniformly sampled transceiver locations and tested on transceiver locations sampled from Gaussian distributions with varying standard deviations.
		(b)~Matching scores between the variables $\mathbfcal{W}$ and $\mathbfcal{V}$ for the proposed UWMMSE in layers $1$,$2$,and $3$ and the classical WMMSE in iterations $1$,$2$,$3$,$50$,and $100$. 
		(c)~Variation in normalized mean sum-rate achieved by UWMMSE and WMMSE under controlled distortion in the input CSI tensor $\mathbfcal{H}$ as compared to the ideal performance achieved on the same CSI tensors without distortion.}
		}
		\vspace{-4mm}
		\label{Fig:ensemble}
\end{figure*}

In this section, we present an empirical convergence analysis of the proposed UWMMSE model and also provide a comparison with the WMMSE algorithm under the same framework.
Our analysis focuses on the variation of the transformed receiver-weight $\mathbfcal{W}$ and its similarity with the final transmitter beamformer output $\mathbfcal{V}$.
Essentially, it is $\mathbfcal{W}$ that represents the channel conditions between a given transceiver pair.
Intuitively, transceivers with better channel conditions should transmit with high power as opposed to transceivers with poor channel conditions, which must hold transmission to conserve power.
Therefore, $\mathbfcal{W}$ plays a key role in driving $\mathbfcal{V}$ to its near-optimal value.
Also, $\mathbfcal{W}$ is a function of $\mathbfcal{V}$ and depends on it to measure the quality of the channel.
To emphasize this inter-dependence which leads to convergence of the UWMMSE model, we extract $\mathbf{W}_i^{(k)}$ and $\mathbf{V}_i^{(k)}$ for all nodes $i \in {1,\dots,M}$ in all layers $k \in {1,2,3}$ for all $10000$ test samples.
Typically, a large value of $\lVert\mathbf{W}_i\rVert_F$ represents a strong channel suitable for transmission and must be allocated greater power. 
On the contrary, a smaller value signifies a poor channel that should not transmit.
To validate this hypothesis, we threshold $\mathbf{W}_i$ for all $i$ such that all transmitters with $\lVert \mathbf{W}_{i}\rVert_F > 1.0$ are assigned a scalar $1.0$ and all transmitters with $\lVert \mathbf{W}_{i}\rVert_F \leq 1.0$ are assigned $0.0$.
Further, we compute the real-valued final power allocation vector $\mathbf{p}$ for the entire network as $\lVert \mathbf{V}_{i}\rVert_F$ for all $i$ and threshold it to yield a binary power allocation vector.
Analogous to binary classification, we treat $\mathbf{p}$ as ground-truth and (the thresholded) $\mathbf{w}$ as the prediction and then compute the F1-score between the two vectors to match them.
The choice of the metric is to ensure that both false-positives and false-negatives have an impact on the score. 
These scores are then averaged across all test samples to find a global trend in matching the two variables.
Similarly, a matching score between $\hat{\mathbfcal{W}},\mathbfcal{V}$ is computed for the case of the WMMSE algorithm.
Since WMMSE takes $100$ iterations to offer best performance, we compute the scores for iterations $1$,$2$, and $3$ for a direct comparison with UWMMSE and also iterations $50$ and $100$ for a more complete comparison.
It is important to note here that we use the respective final power allocations for UWMMSE and WMMSE for this experiment. 
This is because we have already established in Section~\ref{Ss:performance_comparison} that UWMMSE reaches a better convergence than WMMSE in terms of sum-rate and therefore the two algorithms are unlikely to reach an identical final $\mathbf{p}$.
The main objective of this experiment is, however, to evaluate how fast the proposed UWMMSE achieves its respective convergence as compared to the classical WMMSE.
The comparison is shown in Fig~\ref{Fig:convergence}. 
Clearly, UWMMSE offers a better match between $\mathbfcal{W}$ and $\mathbfcal{V}$ in all $3$ layers as compared to the first $3$ iterations of WMMSE. 
Notwithstanding the error bars, we observe that the proposed UWMMSE learns the near-optimal $\mathbfcal{W}$ much earlier as compared to WMMSE. 

\subsection{Robustness}\label{Ss:robust}

We analyze the robustness of the proposed UWMMSE model to distortion in channel state information. 
An analysis of this form is essential since, under real-world scenarios, the channel estimation process is imperfect~\cite{chowdhury2022stability}. 
Yet, it is imperative that the beamforming algorithms are robust to the extent that they are able to maintain a steady performance in spite of these variations.
To that end, we add random Gaussian noise of bounded variance to the channel coefficients in the CSI tensor.
The sum-rate, however, is computed for the undistorted CSI tensor $\mathbfcal{H}$.
For example, in the case of the Rayleigh channel model, the distorted elements $[\bar{\mathbf{H}}_{ij}]_{rt}$ of the input tensor $\bar{\mathbfcal{H}}$ are given by 
%
\begin{align*}
    [\bar{\mathbf{H}}_{ij}]_{rt} &= [\mathbf{H}_{ij}]_{rt} + (c+\sqrt{-1}d) \quad \forall\  r,t.\\
    &\text{where, } c \sim \mathcal{N}(0,\sigma_r), d \sim \mathcal{N}(0,\sigma_r)
\end{align*}
where, $\sigma_r = 0.001$. 
The elements which are distorted are chosen uniformly at random from all the entries of the tensor.
Fig~\ref{Fig:robustness} shows that under a controlled rate of distortion that varies between $0.0$ (no element of $\mathbfcal{H}$ is distorted) to $1.0$ (all elements of $\mathbfcal{H}$ are distorted), the proposed UWMMSE maintains a sum-rate that is better than WMMSE-with-undistorted-input until about $20\%$ of the channel coefficients are distorted. 
Although the performance dips beyond the $20\%$ mark, it is still better than WMMSE-with-distorted-input until about the $60\%$ mark, beyond which the performance of both algorithms applied to the distorted CSI are similar.
Clearly, UWMMSE is reasonably robust insofar as to achieve a superior sum-rate than the classical method in an event of distorted channel estimation until about $40\%$ of the coefficients are estimated correctly.
This degree of robustness, albeit empirical, is inherent to the model since no robustness criterion is enforced at training.
We strongly expect that methods like adversarial training and noise-regularized training will further enhance the robustness of the proposed model.
That analysis, however, is beyond the scope of this work.




\section{Conclusion}\label{S:Conclusion}

We presented UWMMSE, a hybrid algorithm for fast, efficient, and near-optimal beamforming in complex-valued MU-MIMO WANETs by unfolding the iterations of the classical WMMSE algorithm using complex-valued neural models.
The main contribution of this work lies in forming a synergistic combination of an MLP-based parametric functional transformation with a GNN-based learner appropriate for tackling wireless network graphs.
Superiority of this method is established through extensive experiments.
Further, the proposed model is lightweight on account of extensive parameter sharing and also easy to implement in a distributed fashion.
The per-layer computational complexity of UWMMSE matches that of the per-iteration complexity of WMMSE.
However, the main gain lies in the reduction of number of layers in UWMMSE as compared to the number of iterations necessary for WMMSE to converge.
Future work will involve considering time-varying channels with long-term constraints on the beamforming selection such as fairness or battery constraints.
Evaluating the model performance on real-world datasets is also an important next step.


\begin{appendices}
\section{Proof of Proposition I}\label{app:permEq}
\begin{myproof}
    Let $\mathbfcal{V}^{(k)} = \Lambda_k(\mathbfcal{H}, \mathbfcal{V}^{(k-1)}; \Theta)$ denote the output of the UWMMSE architecture in \eqref{E:unfold_u}-\eqref{eq:nonlinear_beta} at layer $k$. 
	Also, $\tilde{\mathbfcal{H}} = \bbPi \mathbfcal{H} \bbPi^\top$ denotes an arbitrary permuted version of the channel tensor with the permutations being enforced only on the first two dimensions of the tensor. 
    and $\tilde{\mathbfcal{V}}^{(k-1)} = \bbPi {\mathbfcal{V}}^{(k-1)}$ represents input $\mathbfcal{V}^{(k-1)}$ to the $k$th layer, with the permutations being enforced on the first dimension of the tensor only.
	Further, $\tilde{\bbS} = \bbPi \bbS \bbPi^\top$ is a permuted version of the transformed CSI matrix, and $\tilde{\bbQ} = \bbPi \bbQ$ is the equivalent permutation of the node feature vectors.
	Firstly, we want to prove that $\Lambda_k(\tilde{\mathbfcal{H}}, \tilde{\mathbfcal{V}}^{(k-1)}; \Theta) = \bbPi \mathbfcal{V}^{(k)}$. 
	We know that $\tilde{\xi}^{(k)} = \Psi(\tilde{\bbS},\tilde{\bbQ}; \theta) = \bbPi {\xi}^{(k)}$ since $\Psi$ is permutation equivariant. 
	Let node $i$ be assigned a new index $\pi(i)$ after permutation $\bbPi$. Then by setting $\tilde{\bbH}_{ij} = [\tilde{\mathbfcal{H}}]_{ij::}$, it follows from \eqref{E:unfold_u} that
	\begin{align*}
	\tilde{\mathbf{U}}^{(k)}_i &\!=\! \bigg(\!\sum_{\substack{j\neq i}}\tilde{\mathbf{H}}_{ij}^{}\tilde{\mathbf{V}}_{j}^{(k-1)}\tilde{\mathbf{V}}_{j}^{(k-1)^H}\!\tilde{\mathbf{H}}_{ij}^{H} \!+ \!\sigma^2\mathbf{I}_R\!\bigg)^{-1}\!\!\!\!\tilde{\mathbf{H}}_{ii}^{}\tilde{\mathbf{V}}_{i}^{(k-1)}\\
	&\! = \! \bigg(\!\sum_{\substack{j\neq i}}{\mathbf{H}}_{\pi(i)\pi(j)}^{}{\mathbf{V}}_{\pi(j)}^{(k-1)}{\mathbf{V}}_{\pi(j)}^{(k-1)^H}\!{\mathbf{H}}_{\pi(i)\pi(j)}^{H} \!+ \!\sigma^2\mathbf{I}_R\!\bigg)^{-1}\\ &\quad\quad\quad\quad\quad\quad\quad\quad {\mathbf{H}}_{\pi(i)\pi(i)}^{}{\mathbf{V}}_{\pi(i)}^{(k-1)}= \mathbf{U}^{(k)}_{\pi(i)},
	\end{align*}
	which, in tensor form, results in $\tilde{\mathbfcal{U}}^{(k)} = \bbPi {\mathbfcal{U}}^{(k)}$. 
	Likewise, from \eqref{E:unfold_w} and \eqref{E:unfold_v} it can be derived that $\tilde{\mathbfcal{W}}^{(k)} = \bbPi {\mathbfcal{W}}^{(k)}$ and $\tilde{\mathbfcal{V}}^{(k)} = \Lambda_k(\tilde{\bbH}, \tilde{\mathbfcal{V}}^{(k-1)}; \Theta)= \bbPi {\mathbfcal{V}}^{(k)}$, as required. 
	
	Leveraging these identities, we now want to show that $\Lambda(\cdot\,; \bbTheta)$ is equivariant. 
	Specifically, for the special case of $K=1$, it can be obtained from the definition of $\Lambda(\cdot; \Theta)$, that
	\begin{align*}
	\Lambda( \tilde{\mathbfcal{H}}; \Theta) &= \Lambda_1(\tilde{\mathbfcal{H}}, \mathbfcal{V}^{(0)}; \Theta) = \Lambda_1(\tilde{\mathbfcal{H}}, \bbPi \mathbfcal{V}^{(0)}; \Theta)\\
	&= \bbPi \Lambda_1({\mathbfcal{H}}, \mathbfcal{V}^{(0)}; \Theta) = \bbPi \Lambda({\mathbfcal{H}}; \Theta),
	\end{align*}
    the fact that $\mathbfcal{V}^{(0)}$ is a constant tensor gives rise to the second equality while the third equality is obtained as a special case of the  previous identity for $k=1$.
	This completes the proof for permutation equivariance of a single-layered UWMMSE. For UWMMSE with $K>1$, permutation equivariance can be established via a simple induction argument omitted here.
\end{myproof}

\section{Proof of Theorem 1}\label{app:thrm}
\begin{myproof}
        This proof is inspired by that of the convergence result presented in~\cite{chowdhury2021unfolding}, and the theoretical linear convergence results of unfolded ISTA~\cite{gregor2010learning}, as presented in~\cite{chen2018theoretical}.
    We assume that $\Tr\left(\mathbf{V}_i^{(k)}\mathbf{V}_i^{{(k)}^H}\right) \to P_i^{*} = \Tr\left(\mathbf{V}_i^{*}\mathbf{V}_i^{{*}^H}\right)$ for all $i$ uniformly. 
    This means that, for all $\eta > 0$, there exists a layer index $K_1$ such that for all $k > K_1$, $\mathbfcal{V}^{(k-1)} = \mathbfcal{V}^* + \mathbfcal{E}_V$ and $\mathbfcal{V}^{(k)} = \mathbfcal{V}^* + \mathbfcal{E}'_V$ where $\lVert [\mathbfcal{E}_V]_i \rVert_F < \eta$ and $\lVert [\mathbfcal{E}_V']_i \rVert_F < \eta$ for all $i$.
	Following notations identical to that of the proof of Proposition~\ref{P:equiv} in Appendix~\ref{app:permEq}, we have that $\mathbfcal{V}^{(k)} = \Lambda_k(\mathbfcal{H}, \mathbfcal{V}^{(k-1)}; \Theta)$. It follows from uniform convergence that,
	\begin{equation}\label{E:proof_necessary_condition}
	\mathbfcal{V}^* + \mathbfcal{E}_V' = \mathbfcal{V}^{(k)} = \Lambda_k(\mathbfcal{H}, \mathbfcal{V}^* + \mathbfcal{E}_V; \Theta).
	\end{equation}
	Therefore, we need to obtain an expression $g_k(\mathbfcal{H}, \mathbfcal{V}^* + \mathbfcal{E}_V; \Theta)$ such that $\Lambda_k(\mathbfcal{H}, \mathbfcal{V}^* + \mathbfcal{E}_V; \Theta) = \mathbfcal{V}^* + g_k(\mathbfcal{H}, \mathbfcal{V}^* + \mathbfcal{E}_V; \Theta)$, that can be replaced in~\eqref{E:proof_necessary_condition} to obtain
	\begin{equation}\label{E:proof_necessary_condition_2}
	\mathbfcal{E}_V' = g_k(\mathbfcal{H}, \mathbfcal{V}^* + \mathbfcal{E}_V; \Theta).
	\end{equation}
	The fact that~\eqref{E:proof_necessary_condition_2} holds for a positive $\eta \to 0$, implies that as $k \to \infty$, $g_k(\mathbfcal{H}, \mathbfcal{V}^* + \mathbfcal{E}_V; \Theta) \to 0$. In what follows, we endeavour to determine the explicit form of $g_k(\mathbfcal{H}, \mathbfcal{V}^* + \mathbfcal{E}_V; \Theta)$ and thereby demonstrate~\eqref{E:convergence_necessary_condition}. 

	First, focusing on an arbitrary $i$, we replace $\bbV_{i}^{(k-1)}$ by $\bbV_i^* + \mathcal{E}_{V_i}$ in~\eqref{E:unfold_v}, where $\mathcal{E}_{V_i} = [\mathbfcal{E}_{V}]_i$  to obtain 
	\begin{align}
	\mathbf{U}^{(k)}_i \!&=\! \bigg(\!\sum_{\substack{j}}\mathbf{H}_{ij}^{}(\mathbf{V}_{j}^{*}+\mathcal{E}_{V_j}){(\mathbf{V}_{j}^{*}+\mathcal{E}_{V_j})}^H\!\mathbf{H}_{ij}^{H} \!+ \!\sigma^2\mathbf{I}_R\!\bigg)^{-1} \nonumber \\
    &\quad\quad\quad\quad\quad\quad\quad\quad \mathbf{H}_{ii}^{}{(\mathbf{V}_{i}^{*}+\mathcal{E}_{V_i})}^H \nonumber
    \end{align}
        {Since we operate in the low-noise regime~\cite{chowdhury2021unfolding}, the noise term in the inverse can be neglected in comparison to the interference term to yield} 
        \begin{align}
        \mathbf{U}^{(k)}_i \!&=\! \bbA_i^{-1}\bbB_i \nonumber
        \end{align}
        where, we define two dummy variables $\bbA_i = \!\sum_{\substack{j}}\mathbf{H}_{ij}^{}(\mathbf{V}_{j}^{*}+\mathcal{E}_{V_j}){(\mathbf{V}_{j}^{*}+\mathcal{E}_{V_j})}^H\!\mathbf{H}_{ij}^{H}$, and $\bbB_i = \mathbf{H}_{ii}^{}{(\mathbf{V}_{i}^{*}+\mathcal{E}_{V_i})}^H$. Defining a third dummy variable $\bbC_i = \sum_{\substack{j}}\mathbf{H}_{ij}^{}\mathbf{V}_{j}^{*}{\mathbf{V}_{j}^{*}}^H\!\mathbf{H}_{ij}^{H}$ we have,   
        \begin{align}
        \mathbf{U}^{(k)}_i \!&=\! \bbA_i^{-1}\bbC_i\bbC_i^{-1}\bbB_i \nonumber \\  
        &= \bbC_i^{-1}\bbB_i + \bbA_i^{-1}(\bbC_i - \bbA_i)\bbC_i^{-1}\bbB_i \nonumber \\ 
        &= \bbC_i^{-1}\hat{\bbB_i} + \bbC_i^{-1}\bbH_{ii}{\mathcal{E}^H_{V_i}} - \bbA_i^{-1}\bigg[\sum_{\substack{j}}\mathbf{H}_{ij}^{}(\mathbf{V}_{j}^{*}{\mathcal{E}^H_{V_j}} \nonumber \\ &\quad\quad\quad\quad + \mathcal{E}_{V_j}{\mathbf{V}_{j}^{*}}^H + \mathcal{E}_{V_j}{\mathcal{E}^H_{V_j}})\mathbf{H}_{ij}^{H}\bigg]\bbC_i^{-1}\bbB_i \tag{where, $\hat{\bbB}_i = \mathbf{H}_{ii}^{}{\mathbf{V}_{i}^{*}}^H$} \nonumber
        \end{align}
        Clearly, the first term represents $\bbU^{*}_i$. All remaining terms that depend on $\mathcal{E}_V$, constitute $\mathcal{E}_U$ such that $||\mathcal{E}_{U_i}||_F \to 0$ as $\eta \to 0$ for all $i$. Thus, $\mathbf{U}^{(k)}_i$ takes the form,
        \begin{align}\label{E:proof_necessary_condition_3}
        \mathbf{U}^{(k)}_i = \mathbf{U}_{i}^* + \mathcal{E}_{U_i}, \,\,\text{\,for all\,} i.
        \end{align}
        We then consider~\eqref{E:unfold_w}, where we replace $\bbU^{(k)}$ and $\bbV^{(k-1)}$, by $\bbU^* + \mathcal{E}_{U}$ and $\bbV^* + \mathcal{E}_{V}$, respectively. Similar to the procedure followed in~\eqref{E:proof_necessary_condition_3}, and leveraging the continuity of $\Phi_{\xi}$, we have
	\begin{align}\label{E:proof_necessary_condition_4}
	w^{(k)}_i = w_i^* + \Phi_{\xi_i^{(k)}}(w^{(k)}_i) + \epsilon_{w_i}, \text{\,for all\,} i 
	\end{align}
	where $|\epsilon_{w_i}| \to 0$ as $\eta \to 0$ for all $i$.
	  Next, we repeat the procedure for~\eqref{E:unfold_v}.
        To that end, we redefine the dummy variables $\bbA_i,\bbB_i$ and $\bbC_i$ as, $\bbA_i = \!\sum_{\substack{j}}\mathbf{H}_{ij}^{H}\mathbf{U}_{j}^{(k)}w_{j}^{(k)}{\mathbf{U}_{j}^{(k)}}^H\!\mathbf{H}_{ij}^{}$, $\bbB_i = \mathbf{H}_{ii}^{H}{\mathbf{U}_{i}^{(k)}}w_i^{(k)}$, and $\bbC_i = \sum_{\substack{j}}\mathbf{H}_{ij}^{H}\mathbf{U}_{j}^{*}w_j^{*}{\mathbf{U}_{j}^{*}}^H\!\mathbf{H}_{ij}^{}$.
        Further, we replace $\bbU_i^{(k)}$ and $w_i^{(k)}$, by their respective forms obtained in~\eqref{E:proof_necessary_condition_3} and~\eqref{E:proof_necessary_condition_4}.
        Clearly, for $k \to \infty$, the non-linearity $\beta(\cdot)$ is no longer significant since $\Tr\left(\bar{\mathbf{V}}_i^{(k)}\bar{\mathbf{V}}_i^{{(k)}^H}\right) \to P_i^*$ and $0< P_i^* < P_{\max}$ for all $i$. 
	Therefore, from~\eqref{E:unfold_v}-\eqref{eq:nonlinear_beta} we get that, 
\begin{align}\label{E:proof_necessary_condition_5}
	\bbV^{(k)}_i &= \bbC_i^{-1}\bbB_i + \bbA_i^{-1}(\bbC_i - \bbA_i)\bbC_i^{-1}\bbB_i \nonumber \\
        &= \bbC_i^{-1}\hat{\bbB}_i + \mathcal{E}_{V_i} + \bbA_i^{-1}\bigg[\bbA_i^{}\Phi_{\xi_{i}^{(k)}}(w_i^{(k)})\bigg]\bbC_i^{-1}\bar{\bbB}_i \nonumber \\ &\quad\quad + \bbA_i^{-1}\bigg[(\bbC_i - \bbA_i)(w_i^{*}+\Phi_{\xi_{i}^{(k)}}(w_i^{(k)}))\bigg]\bbC_i^{-1}\bar{\bbB}_i \tag{where, $\hat{\bbB}_i = \mathbf{H}_{ii}^{H}{\mathbf{U}_{i}^{*}}w_{i}^*$ and $\bar{\bbB}_i = \mathbf{H}_{ii}^{H}{\mathbf{U}_{i}^{*}}$},  \nonumber \\
        &= \bbC_i^{-1}\hat{\bbB}_i + \mathcal{E}_{V_i} + \bbA_i^{-1}\bigg[\sum_{\substack{j\neq i}}\mathbf{H}_{ij}^{H}\mathbf{U}_{j}^{(k)}(w_j^* \Phi_{\xi_{i}^{(k)}}(w_i^{(k)}) \nonumber \\ &\quad\quad - w_i^* \Phi_{\xi_{j}^{(k)}}(w_j^{(k)})){\mathbf{U}_{j}^{(k)}}^H\!\mathbf{H}_{ij}^{}\bigg]\bbC_i^{-1}\bar{\bbB}_i \text{\,for all\,} i \label{E:proof_necessary_condition_5}
        \end{align}
	where $||\mathcal{E}_{V_i}||_F \to 0$ as $\eta \to 0$ for all $i$.
	Since, $(\mathbfcal{U}^*, \mathbfcal{W}^*, \mathbfcal{V}^*)$ form a fixed point of the WMMSE updates~\cite{shi2011iteratively}, $\bbV_i^*$ is given by the first term in the r.h.s of~\eqref{E:proof_necessary_condition_5} for all $i$. 
	Therefore, the last two terms in the r.h.s of~\eqref{E:proof_necessary_condition_5} constitute, element-wise, the function $g_k(\mathbfcal{H}, \mathbfcal{V}^* + \mathbfcal{E}; \Theta)$. 
	Further, since $||\mathcal{E}_{V_i}||_F \to 0$ as $\eta \to 0$ (equivalently, as $k \to \infty$), as $k$ goes to infinity, the last term in~\eqref{E:proof_necessary_condition_5} must go to $0$, thus completing the proof. 
\end{myproof}

\end{appendices}


\bibliographystyle{IEEEtran}
\bibliography{citations}

\end{document}

%% file: my_sections.tex
\usepackage{needspace}



%% file: figures/tikz_styles.tex
\tikzstyle{phantom vertex} = [ ellipse, 
                               anchor = center, 
                               minimum height = 1*\unit, 
                               minimum width  = 1*\unit,
                               inner sep=0pt,
                               anchor=center]
\tikzstyle{red vertex}   = [black, fill = red!20,   phantom vertex, draw]
\tikzstyle{black vertex} = [black, fill = black!20, phantom vertex, draw]
\tikzstyle{blue vertex}  = [black, fill = blue!20,  phantom vertex, draw]
\tikzstyle{green vertex} = [black, fill = green!20,  phantom vertex, draw]
\tikzstyle{yellow vertex} = [black, fill = yellow!20,  phantom vertex, draw]
\tikzstyle{cyan vertex} = [black, fill = cyan!20,  phantom vertex, draw]
\tikzstyle{vertex}       = [draw, phantom vertex]

\tikzstyle{point} = [ellipse, inner sep=0pt, draw, fill=white, anchor = center,
                     minimum height = 0.05*\unit, minimum width  = 0.05*\unit]